\documentclass[useAMS,usenatbib]{mn2e}
\usepackage{graphicx}
\usepackage{rotating}
\usepackage{lscape}
\DeclareMathAlphabet{\mathpzc}{OT1}{pzc}{m}{it}

\def\lsim{\,\lower2truept\hbox{${<\atop\hbox{\raise4truept\hbox{$\sim$}}}$}\,}
\def\gsim{\,\lower2truept\hbox{${> \atop\hbox{\raise4truept\hbox{$\sim$}}}$}\,}

\title[Mid-/far-IR spectroscopic surveys]{Exploring the early dust-obscured phase of galaxy formation with blind mid-/far-IR spectroscopic surveys 
}
\author[M. Bonato et al.]
{M. Bonato$^{1,2}$\thanks{matteo.bonato@oapd.inaf.it},
M. Negrello$^{2}$,
Z.-Y. Cai$^{3}$,
G. De Zotti$^{2,3}$,
A. Bressan$^{3}$,
A. Lapi$^{3,4}$,
\newauthor
C. Gruppioni$^{5}$,
L. Spinoglio$^{6}$ and
L. Danese$^{3}$ \\
$^{1}$Dipartimento di Fisica e Astronomia ``G.Galilei'', Universit\`a degli Studi di Padova, Vicolo Osservatorio 3, I-35122 Padova, Italy \\
$^{2}$INAF, Osservatorio Astronomico di Padova, Vicolo Osservatorio 5, I-35122 Padova, Italy \\
$^{3}$SISSA, Via Bonomea 265, I-34136 Trieste, Italy \\
$^{4}$Dipartimento di Fisica, Universit\`a ``Tor Vergata'', Via della Ricerca Scientifica 1, I-00133 Roma, Italy \\
$^{5}$INAF, Osservatorio Astronomico di Bologna, Via Ranzani 1, I-40127 Bologna, Italy \\
$^{6}$Istituto di Astrofisica e Planetologia Spaziali, INAF-IAPS, Via Fosso del Cavaliere 100, I-00133 Roma, Italy}
\date{Released 2013 Xxxxx XX}

\pagerange{\pageref{firstpage}--\pageref{lastpage}} \pubyear{2013}

\def\LaTeX{L\kern-.36em\raise.3ex\hbox{a}\kern-.15em
    T\kern-.1667em\lower.7ex\hbox{E}\kern-.125emX}
\def\simlt{\mathrel{\rlap{\lower 3pt\hbox{$\sim$}}\raise 2.0pt\hbox{$<$}}}
\def\simgt{\mathrel{\rlap{\lower 3pt\hbox{$\sim$}}\raise 2.0pt\hbox{$>$}}}

\begin{document}

\label{firstpage}

\maketitle

\begin{abstract}
While continuum imaging data at far-infrared to sub-millimeter wavelengths have provided tight constraints on the population  properties of dusty star forming galaxies up to high redshifts, future space missions like the Space Infra-Red Telescope for Cosmology and Astrophysics (SPICA) and ground based facilities like the Cerro Chajnantor Atacama Telescope (CCAT) will allow  detailed investigations of their physical properties via their mid-/far-infrared line emission. We present updated predictions for the number counts and the redshift distributions of star forming galaxies spectroscopically detectable by these future missions. These predictions exploit a recent upgrade of evolutionary models, that include the effect of strong gravitational lensing, in the light of the most recent \textit{Herschel} and South Pole Telescope data. Moreover the relations between line and continuum infrared luminosity are re-assessed, considering also differences among source populations, with the support of extensive simulations that take into account dust obscuration. The derived line luminosity functions are found to be highly sensitive to the spread of the line to continuum luminosity ratios. Estimates of the expected numbers of detections per spectral line by SPICA/SAFARI and by CCAT surveys for different integration times per field of view at fixed total observing time are presented. Comparing with the earlier estimates by \citet{Spin12} we find, in the case of SPICA/SAFARI, differences within a factor of two in most cases, but occasionally much larger. More substantial differences are found for CCAT.

\end{abstract}


\section{Introduction}\label{sect:intro}

The rest-frame mid- to far-infrared (IR) spectral region offers a rich suite of spectral lines that allow us to probe all phases of the interstellar medium (ISM): ionized, atomic and molecular \citep{Spin92}. Measurements of these lines provide redshifts and key insight on physical conditions of dust obscured regions and on the energy sources controlling their temperature and pressure. This information is critically important for investigating the complex physics ruling the dust-enshrouded active star forming phase of galaxy evolution and the relationship with nuclear activity.

A major progress in this field is therefore expected with planned or forthcoming projects specifically devoted to mid- to far-IR spectroscopy such as the SPace IR telescope for Cosmology and Astrophysics (SPICA)\footnote{http://www.ir.isas.jaxa.jp/SPICA/SPICA\_HP/index-en.html} with its SpicA FAR infrared Instrument \citep[SAFARI;][]{Roelfsema12} and the Cerro Chajnantor Atacama Telescope \citep[CCAT;][]{Woody2012}\footnote{http://www.ccatobservatory.org}. SAFARI is an imaging spectrometer designed to fully exploit the extremely low far-IR background environment provided by the SPICA observatory, whose telescope will be actively cooled at 6K. In each integration it will take complete 34--210\,$\mu$m spectra with three bands ($\,34-60\,\mu$m, $\,60-110\,\mu$m, $110-210\,\mu$m), spatially resolving the full 2$^{\prime}\times$2$^{\prime}$ field of view (FoV). CCAT is a 25 meter diameter Ritchey-Chretien (RC) telescope, that supports cameras and spectrometers operating in the 0.2--2.1 mm wavelength range with a goal FoV of $1\,\hbox{deg}^2$ (requirement: 20$^{\prime}\times$20$^{\prime}$). 

%
%
\begin{figure*}
\makebox[\textwidth][c]{
\includegraphics[trim=0.8cm 0.6cm 1.5cm 1.2cm,clip=true,width=0.9\textwidth, angle=0]{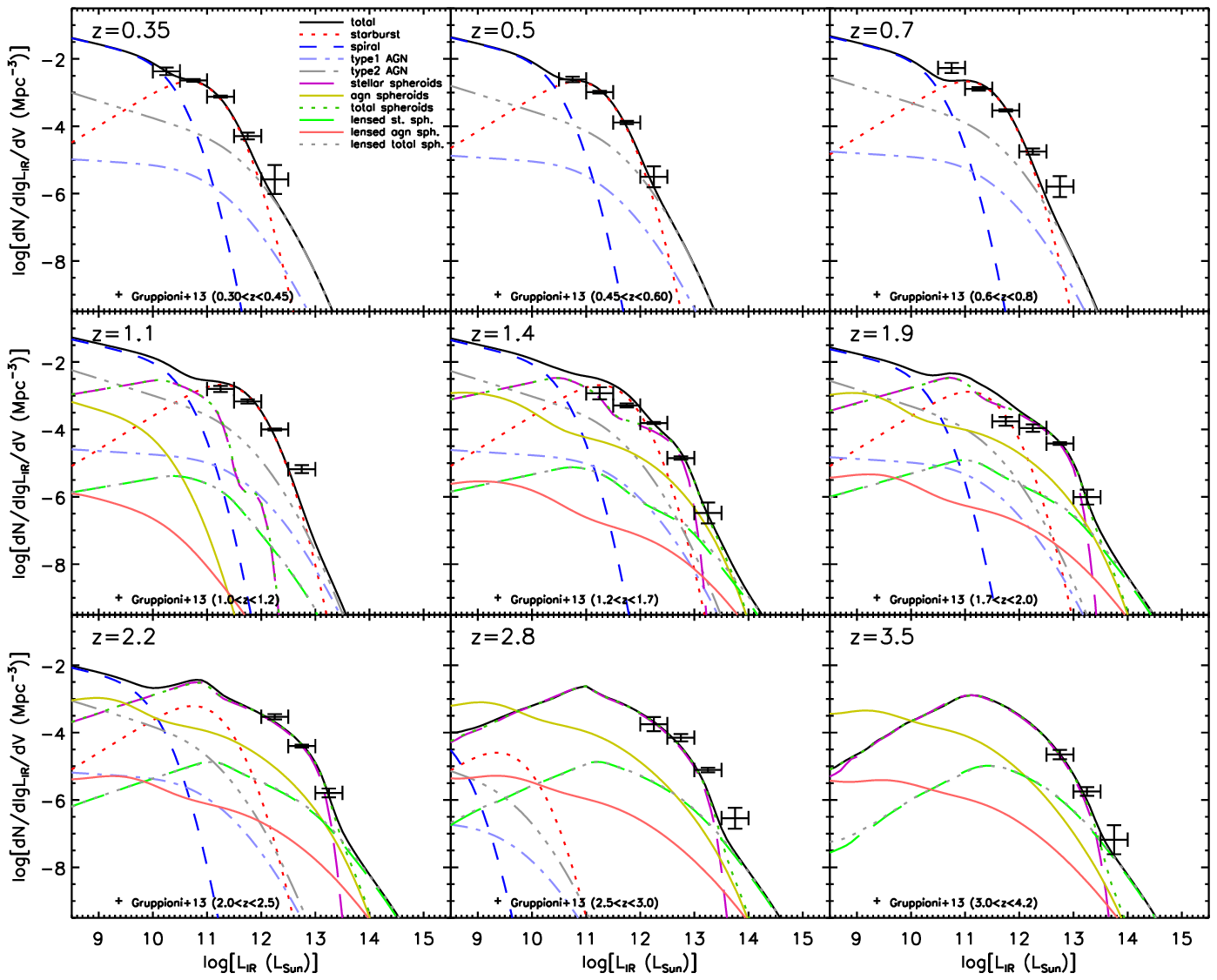}
}
\vspace{-0.3cm}
\caption{Comparison of the IR (8-1000\,$\mu$m) luminosity functions at several redshifts yielded by the \citet{Cai13} model with the observational determinations by \citet{Grupp13}, that became available only after the completion of the \citet{Cai13} paper. At $z\leq 1.5$ the dominant contributions come from ``warm'' (\emph{dotted red lines}) and ``cold'' (\emph{dashed blue lines}) star forming galaxies. Type-2 AGNs (\emph{dash-dot-dotted grey lines}) dominate at the highest IR luminosities while type-1 AGNs (\emph{dash-dotted light blue lines}) are always sub-dominant (in the IR). At $z>1$ we have also contributions from proto-spheroidal galaxies (\emph{long dashed purple lines}) and from the associated AGNs (both obscured and unobscured; \emph{solid yellow lines}). The \emph{dotted dark green lines} (that are generally superimposed to the \emph{long dashed purple lines}) are the combination of the two components. The \emph{long dashed green lines} show the contribution of the lensed stellar proto-spheroidal galaxies to the observed luminosity functions; the contribution of the associated AGNs is shown by the \emph{solid orange lines}, while the \emph{dotted dark grey lines} represent the combination of the two components. These lensed components are computed following \citet{Lapi12}.}
 \label{fig:LF_IR_C13}
\end{figure*}
%
\begin{figure*}
\begin{center}
\includegraphics[width=0.3\textwidth]{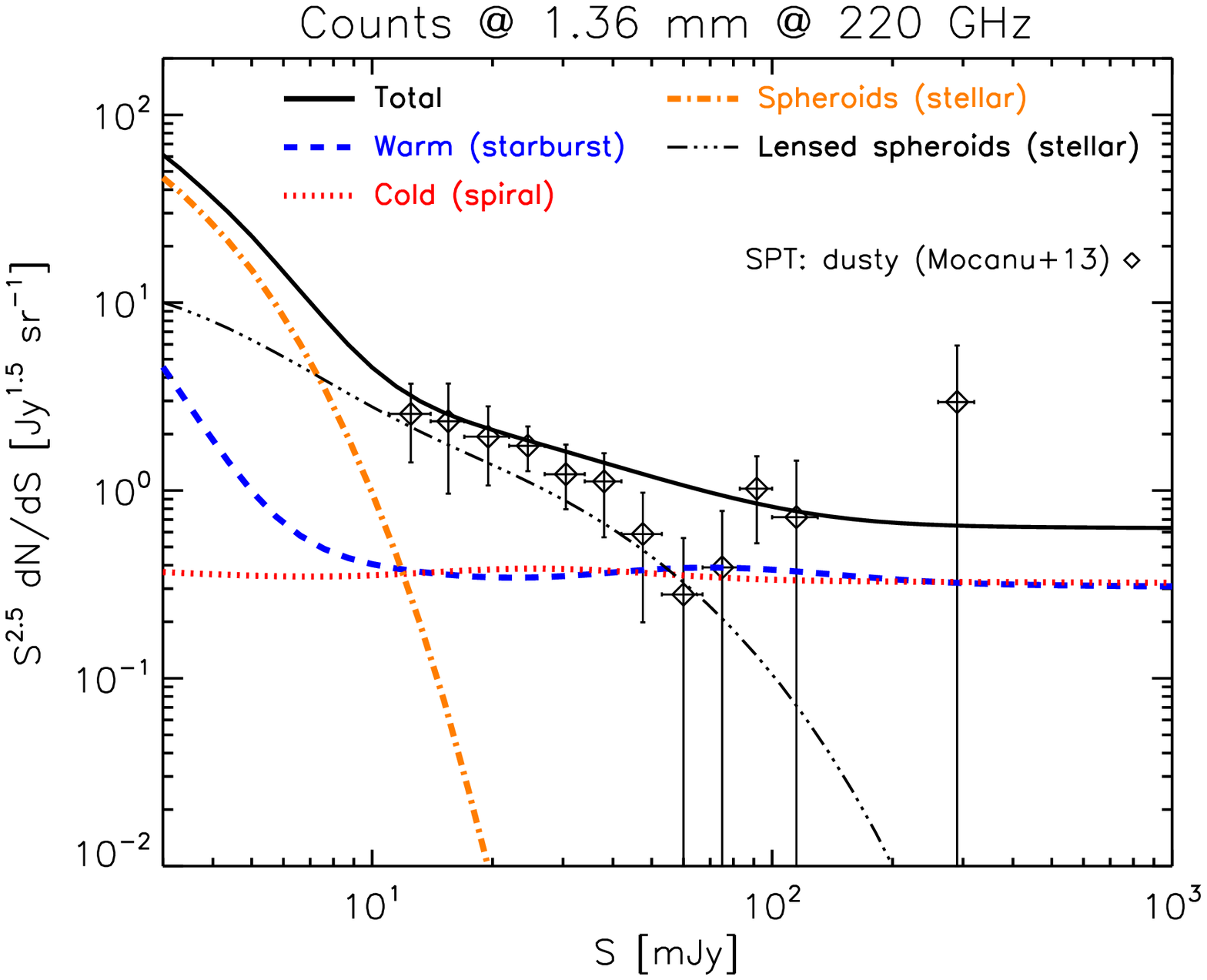} 
\includegraphics[width=0.3\textwidth]{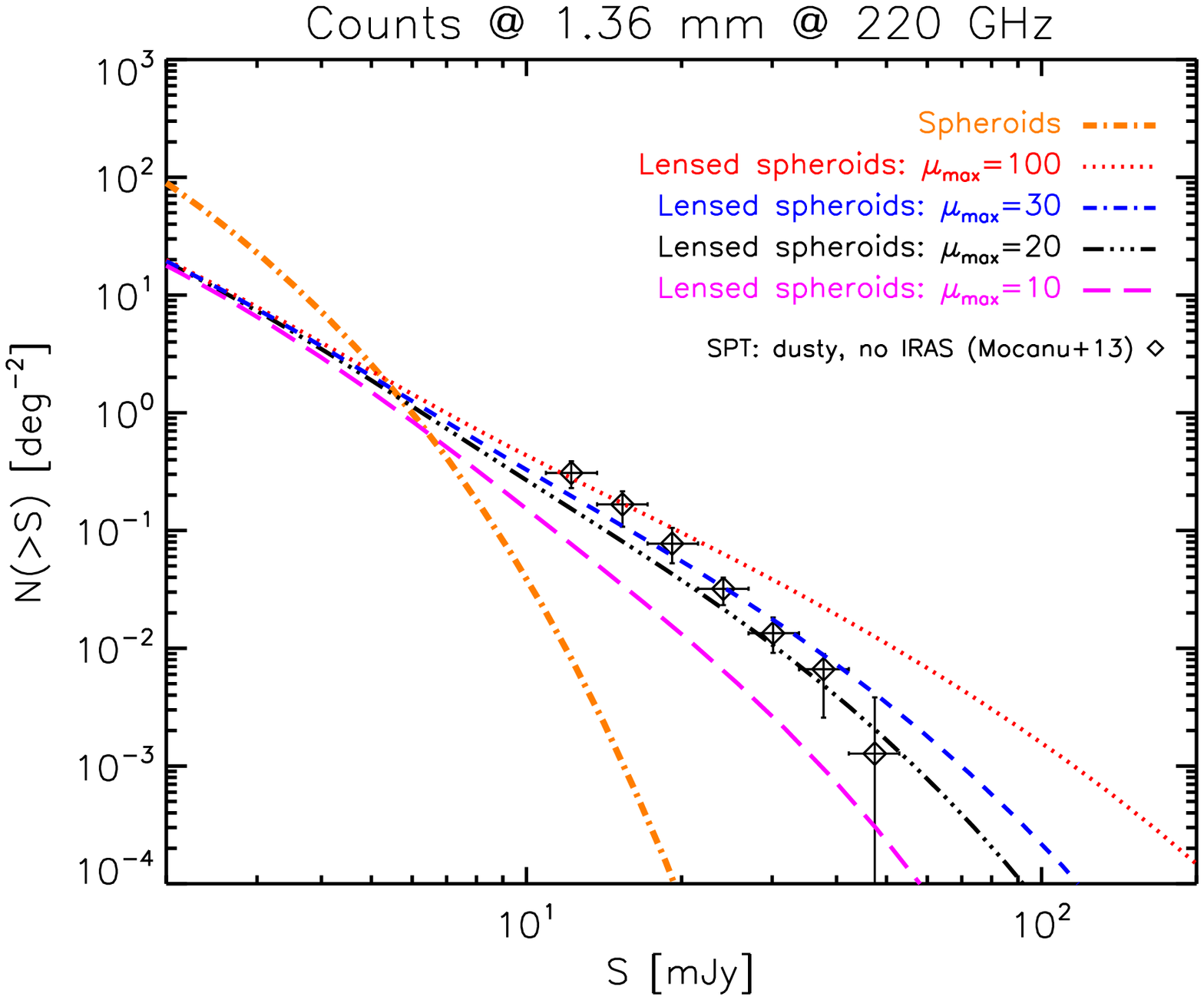} 
\includegraphics[width=0.3\textwidth]{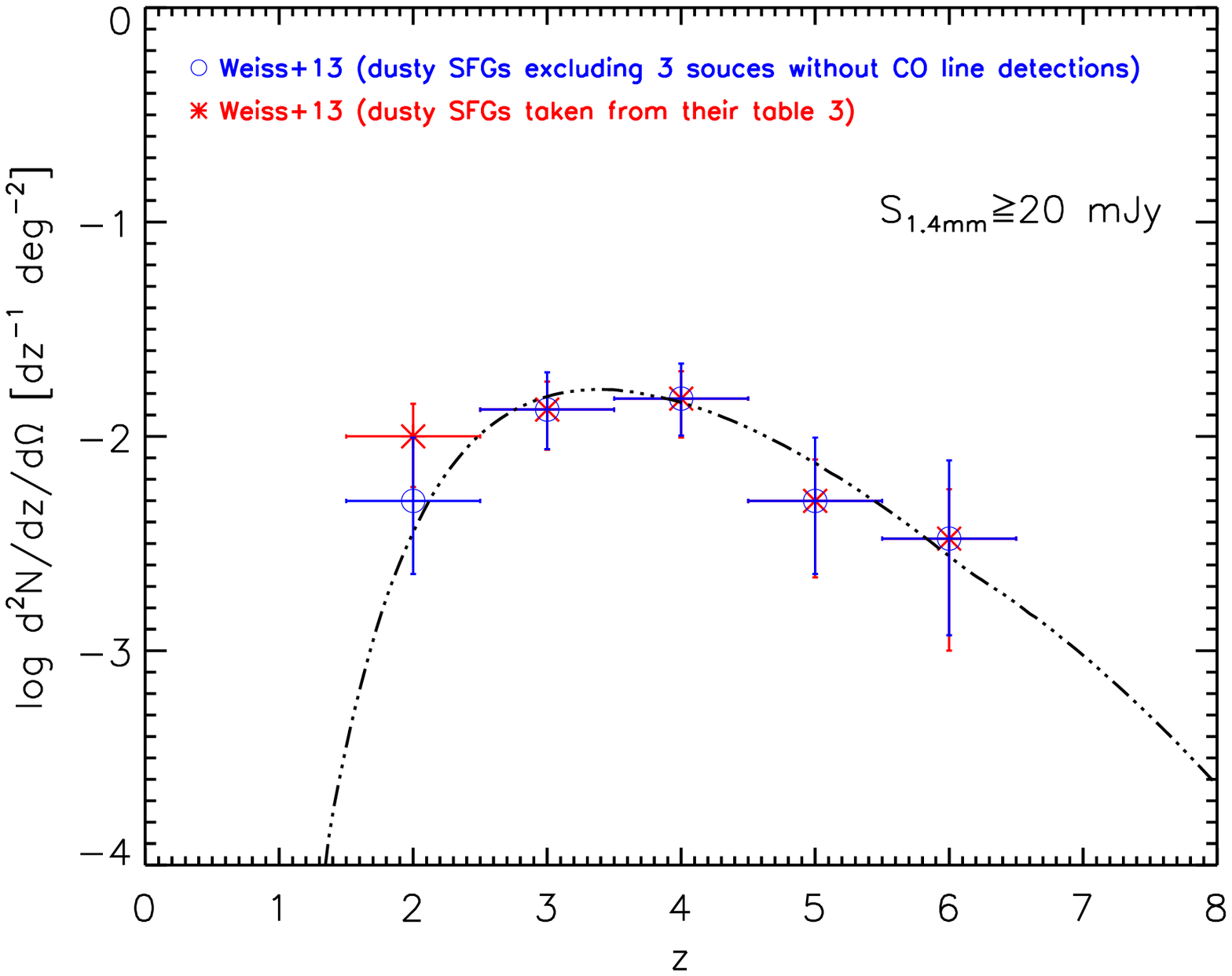} 
 \caption{{\it Left panel}: Euclidean normalized differential SPT counts at 1.36\,mm \citep[220 GHz;][]{Mocanu2013} of all dusty galaxies compared with the prediction of the \citet{Cai13} model. The different lines correspond to the contributions of different source populations, as specified in the legend inside the panel. At variance with the counts presented in the \citet{Cai13} paper, which do not allow for an upper limit to the gravitational amplification (i.e. implicitly assume point-like background sources), here a maximum amplification $\mu_{\rm max}=20$ \citep[corresponding to a source size of $\simeq 3\,$kpc;][]{Lapi12} has been adopted for strongly lensed galaxies.  A comparison between the model and the whole observed counts (not shown here to avoid over-crowding the figure) can be found in \citet{Cai13}. {\it Central panel}:  integral counts of dusty galaxies without counterpart in the IRAS catalogue, interpreted as candidate strongly lensed galaxies, compared with the prediction of the \citet{Cai13} model for different values of $\mu_{\rm max}$. The best match is obtained for $\mu_{\rm max}$ in the range 20--30.  {\it Right panel}: comparison of the redshift distribution of strongly lensed SPT galaxies selected at 1.4\,mm \citep{Weiss13} with predictions of the \citet{Cai13} model. The photometric redshifts of the 3 galaxies without CO line detection are $>3$.}
 \label{fig:dNdF_dNdz_mm_C13}
 \end{center}
\end{figure*}
%

%

Predictions for SPICA-SAFARI and CCAT spectroscopic surveys have been worked out by \citet{Spin12} using phenomenological models for the cosmological evolution of the IR luminosity of galaxies and Active Galactic Nuclei (AGNs) and empirical correlations between line and continuum luminosities. Although this study is quite recent, it is now possible to substantially improve the analysis taking advantage of the many data on the redshift dependent IR luminosity functions of different populations of extragalactic sources that were not available when the evolutionary models used by \citet{Spin12} were built, but are accounted for by the more recent \citet{Cai13} model adopted in this work. Moreover, we have carefully revisited the correlations between line and continuum emissions. Major updates include:
\begin{itemize}
\item An assessment of the effect of the dispersion of line/continuum luminosity ratios, $\sigma_{\ell,c}$. As shown by \citet{Spin12}, at high redshifts SAFARI and CCAT surveys are expected to detect sources in the high luminosity tail of the IR luminosity function. Since the latter is declining very steeply the derived line luminosity functions are strongly sensitive to $\sigma_{\ell,c}$.
\item An analysis of the effect of dust obscuration. Molecular clouds inside which new stars are formed typically have very high optical depths \citep{Silva98,Vega08} implying strong attenuation also at mid-IR wavelengths of lines produced within them. This may bias observations of line/continuum luminosity ratios if, e.g., they preferentially refer to objects with relatively low attenuations.
\item The consideration of the observational evidences that different galaxy populations have different line/continuum luminosity ratios. For example, some IR lines of low-z Ultra-Luminous IR Galaxies (ULIRGs) are found to have much lower equivalent widths than observed for both less luminous galaxies at similar redshifts and for galaxies of similar luminosity at high-$z$.
\item The explicit predictions for the counts of strongly lensed galaxies \citep[e.g.][]{Neg07,Neg10}, that allow us to pierce the properties of galaxies that would otherwise be beyond the detection limits.
\end{itemize}
On the other hand, \citet{Spin12} also gave predictions for lines produced by AGNs. The models used by them either treat starburst galaxies and AGNs as separate populations or adopt simple, only partly validated recipes to estimate the AGN fractional contributions to either the IR source population or to the total IR luminosity. To substantially improve over the \citet{Spin12} results we need to work out a major upgrade of the model to make it capable of dealing in a self-consistent way with both the starburst and the AGN component of IR galaxies. The upgraded model will take into account the current constraints on the distribution of AGN luminosities as a function of the starburst luminosity as well as the obscuration of the nucleus  both by the circum-nuclear torus and by the interstellar dust. The latter may be large during the phases of most intense star formation activity. Work in this direction is currently in progress and the results will be reported in due time (Bo\-na\-to et al., in preparation). We anticipate that the presence of an AGN mostly affects the higher ionisation lines ([NeIII]$15.5\mu$m, [OIV]$26\mu$m, [OIII]$52\mu$m and $88\mu$m and [NIII]$57\mu$m) which can be excited in AGN Narrow Line Regions (NLR) with typical conditions \citep[e.g. densities $N_{\rm H}\sim 10^3$--$10^4\,\hbox{cm}^{-3}$, ionisation potential $\log U$ from $\sim -2$ to $\sim -1$;][]{Spin12}. Because these lines are also excited in HII regions, in a composite object, which has both a star formation and an AGN component, the total line emission will be the sum of the two components. So we expect an increase in the total line emission. The line which will be much brighter (up to a factor of 10) in AGNs with respect to starbursts is the [OIV]$26\mu$m line. Moreover the two lines of [NeV] at 14.3 and $24.3\,\mu$m can be almost exclusively excited by AGNs and are considered AGN spectral signatures  \citep[see, e.g.,][]{Tommasin2010}. The analysis of the emission line spectrum will reveal the nature of the emitting object, i.e. if it is AGN dominated or starburst dominated.

%

The plan of the paper is the following. In Section\,\ref{sect:evol} we present the adopted model \citep{Cai13} for the evolution with cosmic time of the  IR (8-1000\,$\mu$m) luminosity function. Section\,\ref{sect:IRlines} contains a brief summary of the information provided by IR lines on the physical conditions of the emitting regions. In Section\,\ref{sect:line_vs_IR} we discuss the line to IR luminosity ratios for the main mid/far-IR lines. In Section\,\ref{sect:LF} we work out our predictions for line luminosity functions, number counts and redshift distributions in the SPICA/SAFARI and CCAT bands. In Section\,\ref{sect:survey} we discuss possible SPICA/SAFARI and CCAT survey strategies, considering different integration times per FoV and areal coverages. Section\,\ref{sect:conclusions} contains a summary of our main conclusions.


We adopt a flat $\Lambda \rm CDM$ cosmology with matter density $\Omega_{\rm m} = 0.32$, dark energy density $\Omega_{\Lambda} = 0.68$ and Hubble constant $h=H_0/100\, \rm km\,s^{-1}\,Mpc^{-1} = 0.67$ \citep{PlanckCollaborationXVI2013}.

\section{Evolution of the IR luminosity functions}\label{sect:evol}

As mentioned in Sect.~\ref{sect:intro} our reference model for the evolution of the IR luminosity functions is the one recently worked out by \citet{Cai13} based on a comprehensive ``hybrid'' approach. The model starts from the consideration of the observed dichotomy in the ages of stellar populations of early-type galaxies on one side and late-type galaxies on the other \citep[cf.][ their Fig.~10]{Bernardi2010}. Early-type galaxies and massive bulges of Sa galaxies are composed of relatively old stellar populations with mass-weighted ages $\gsim 8$--9\,Gyr (corresponding to formation redshifts $z\gsim 1$--1.5), while the disc components of spirals and the irregular galaxies are characterized by significantly younger stellar populations. Thus the progenitors of early-type galaxies, protospheroids, are the dominant star forming population at $z\gsim 1.5$ (possible examples of such objects are the Herschel sources discussed in \citealt{Ivison13} and \citealt{Fu13}), while IR galaxies at $z\lsim 1.5$ are mostly late-type ``cold'' (normal) and ``warm'' (starburst) galaxies.

The physical model for proto-spheroidal galaxies is based on the approach by \citet[][but see also Lapi et al. 2006, 2011 and Mao et al. 2007]{Granato2004} which hinges upon high resolution numerical simulations showing that dark matter halos form in two stages \citep{Zhao2003,Wang11,LC11}. An early fast collapse of the halo bulk, including a few major merger events, reshuffles the gravitational potential and causes the dark matter and stellar components to undergo (incomplete) dynamical relaxation. A slow growth of the halo outskirts in the form of many minor mergers and diffuse accretion follows. This second stage has little effect on the inner potential well where the visible galaxy resides.

The star formation and the growth of the central super-massive black hole are triggered by the fast collapse/merger phase of the halo and are controlled by self-regulated baryonic processes. They are driven by the rapid cooling of the gas and by the loss of angular momentum by radiation drag, are regulated by the energy feedback from supernovae (SNe) and from the active nucleus and are quenched by the AGN feedback. The latter is relevant especially in the most massive galaxies and is responsible for the shorter duration
(0.5--0.7\,Gyr) of their active star forming phase. In less massive proto-spheroidal galaxies the star formation rate is mostly regulated by SN feedback and continues for a few Gyr. Only a minor fraction of the gas initially associated to the dark matter halo is converted into stars. The rest is ejected by feedback processes. The metal enrichment and the dust formation are rapid (timescales $\sim \hbox{few}\times 10^{7}$\,yr) so that most of the active star formation and black hole growth phase is dust enshrouded. The equations governing the evolution of the baryonic matter in dark matter halos and the adopted values for the parameters are given in the Appendix of the \citet{Cai13} paper where some examples of the evolution with galactic age of quantities related to the stellar and to the AGN component are also shown. Since spheroidal galaxies are observed to be in passive evolution at $z\lsim 1$-1.5 \citep[e.g.,][]{Renzini2006} they are bright at sub-mm wavelengths only at higher redshifts.

This scenario provides a physical explanation for the observed positive evolution of both galaxies and AGNs up to $z\approx2.5$ and for the negative evolution at higher redshifts, for the sharp transition from Euclidean to extremely steep counts at (sub-)mm wavelengths, as well as for the (sub-)mm counts of strongly lensed galaxies, that are hard to account for by alternative, physical or phenomenological, approaches. The model successfully predicted \citep{Neg07} the mm \citep{Vieira10} and sub-mm \citep{Neg10} counts of strongly gravitationally lensed galaxies. Furthermore, as shown by \citet{Xia2012} and \citet{Cai13}, the halo masses inferred from both the angular correlation function of detected sub-mm galaxies \citep{Cooray10,Maddox2010} and from the power spectrum of fluctuations of the cosmic infrared background \citep{Amblard11,PlanckCollaboration2011,Viero13} are fully consistent with those implied by this scenario while are larger than those implied by the major mergers plus top-heavy initial stellar mass function \citep{Baugh05,Lacey10} and smaller than those implied by cold flow models \citep{Dave10}.

The evolution of late-type galaxies and of $z< 1.5$ AGNs is described using a parametric phenomenological approach. For the IR luminosity functions of both ``warm'' starburst galaxies and ``cold'' (normal) late-type galaxies the functional form:
\begin{eqnarray}\label{eq:func_form}
\Phi(L_{\rm IR},z)\,{\rm d}\log L_{\rm IR} = ~~~~~~~~~~~~~~~~~~~~~~~~~~~~~~~~~~~~~~ \nonumber \\
~~~~~~~~\Phi^{\star} \left( \frac{L_{\rm IR}}{L^{\star}} \right)^{1-\alpha}
\exp \left[  -\frac{\log^{2}(1 + L_{\rm IR}/L^{\star})}{2\sigma^{2}}\,{\rm d}\log L_{\rm IR}
\right]
 \end{eqnarray}
advocated by \citet{Saunders1990} was adopted. For the ``warm'' population power law density and luminosity evolution [$\Phi^{\star}(z)=\Phi^{\star}(z=0)\times(1+z)^{\alpha_{\Phi}}$ ; $L^{\star}(z)=L^{\star}(z=0)\times(1+z)^{\alpha_{L}}$]
up to $z_{\rm break}=1$ was assumed. The ``cold'' population, comprising normal disc galaxies, has only a mild luminosity evolution up to the same value of $z_{\rm break}$, as indicated by chemo/spectrophotometric evolution models.  At $z>z_{\rm break}$ both $\Phi^{\star}(z)$ and $L^{\star}(z)$ are kept to the values at $z_{\rm break}$ multiplied by a smooth cut-off function. 
For further details and the values of the parameters, see \citet{Cai13}.

The model accurately fits a broad variety of data\footnote{See figures in http://people.sissa.it/$\sim$zcai/galaxy\_agn/.}: multi-frequency and multi-epoch luminosity functions of galaxies and AGNs, redshift distributions, number counts (total and per redshift bins). Figure\,\ref{fig:LF_IR_C13} illustrates how the model predictions, without any adjustment to take into account the new data, compare to the multi-epoch IR (8-1000\,$\mu$m) luminosity functions observationally estimated by \cite{Grupp13}, that became available only after the completion of the \citet{Cai13} paper. Moreover, the model accurately accounts for the recently determined counts and redshift distribution of strongly lensed galaxies detected by the South Pole Telescope \citep[SPT;][]{Mocanu2013,Weiss13}, also published after the paper was completed (Fig.\,\ref{fig:dNdF_dNdz_mm_C13}). The maximum gravitational amplification, $\mu_{\rm max}$, depends on the source size \citep[see, e.g.,][their Fig.~8]{Lapi12}. The original \citet{Cai13} estimates assumed point like sources. Figure\,\ref{fig:dNdF_dNdz_mm_C13} shows that SPT data constrain $\mu_{\rm max}$ to be $\simeq 20$--30, corresponding to a half stellar mass radius of the source $R_e\simeq 3\,$kpc. A lower value of $\mu_{\rm max}$ ($\simeq 10$) is indicated by the counts of strongly lensed galaxies from $95\,\hbox{deg}^2$ of the \textit{Herschel} Multi-tiered Extragalactic Survey \citep[HerMES;][]{Wardlow2013}. On the other hand, estimated counts from the \textit{Herschel} Astrophysical Terahertz Large Area Survey (H-ATLAS) Science Demonstration Phase (SDP) field \citep{GonzalezNuevo2012} are consistent with the values of $\mu_{\rm max}$ implied by the SPT data. For completeness we also mention that the model over-predicts the $870\,\mu$m counts from the ALMA survey of submillimetre galaxies in the Extended Chandra Deep Field South \citep[ECDFS;][]{Karim2013}. It should be noted, however, that sub-millimeter number counts in the ECDFS are significantly lower  compared to any other deep fields observed at $850\,\mu$m, in line with results from optical/NIR surveys which revealed that several rest-frame optical populations are under-abundant in the CDFS compared to other deep fields \citep{Weiss2009}. All these (and many other) comparisons between model and data are illustrated by figures in the Web site http://people.sissa.it/$\sim$zcai/galaxy\_agn/.
%
\begin{figure*}
\begin{center}
\includegraphics[trim=3.45cm 1.1cm 1.4cm 1.2cm,clip=true,width=0.45\textwidth]{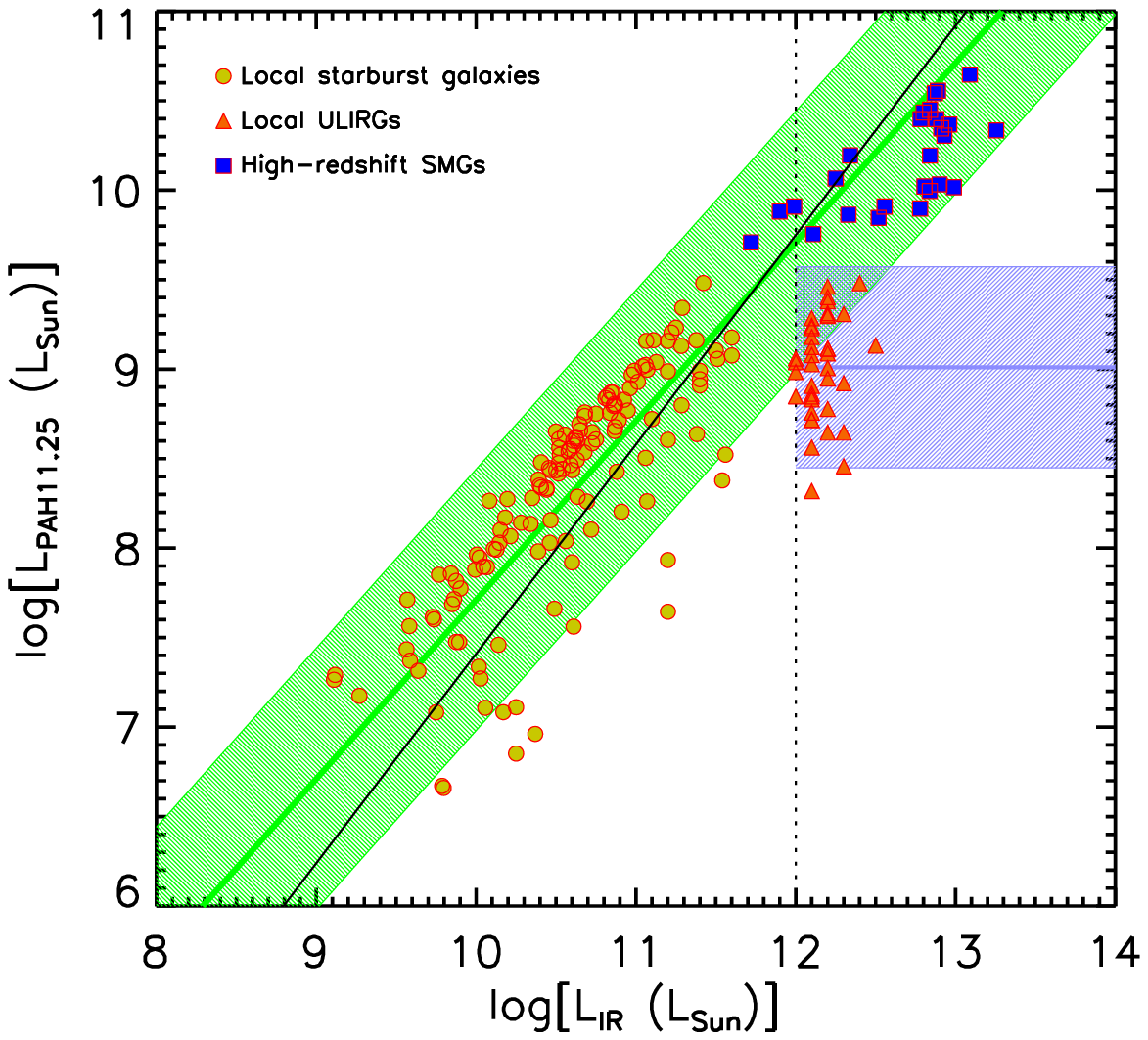}
\includegraphics[trim=3.45cm 1.1cm 1.4cm 1.2cm,clip=true,width=0.45\textwidth]{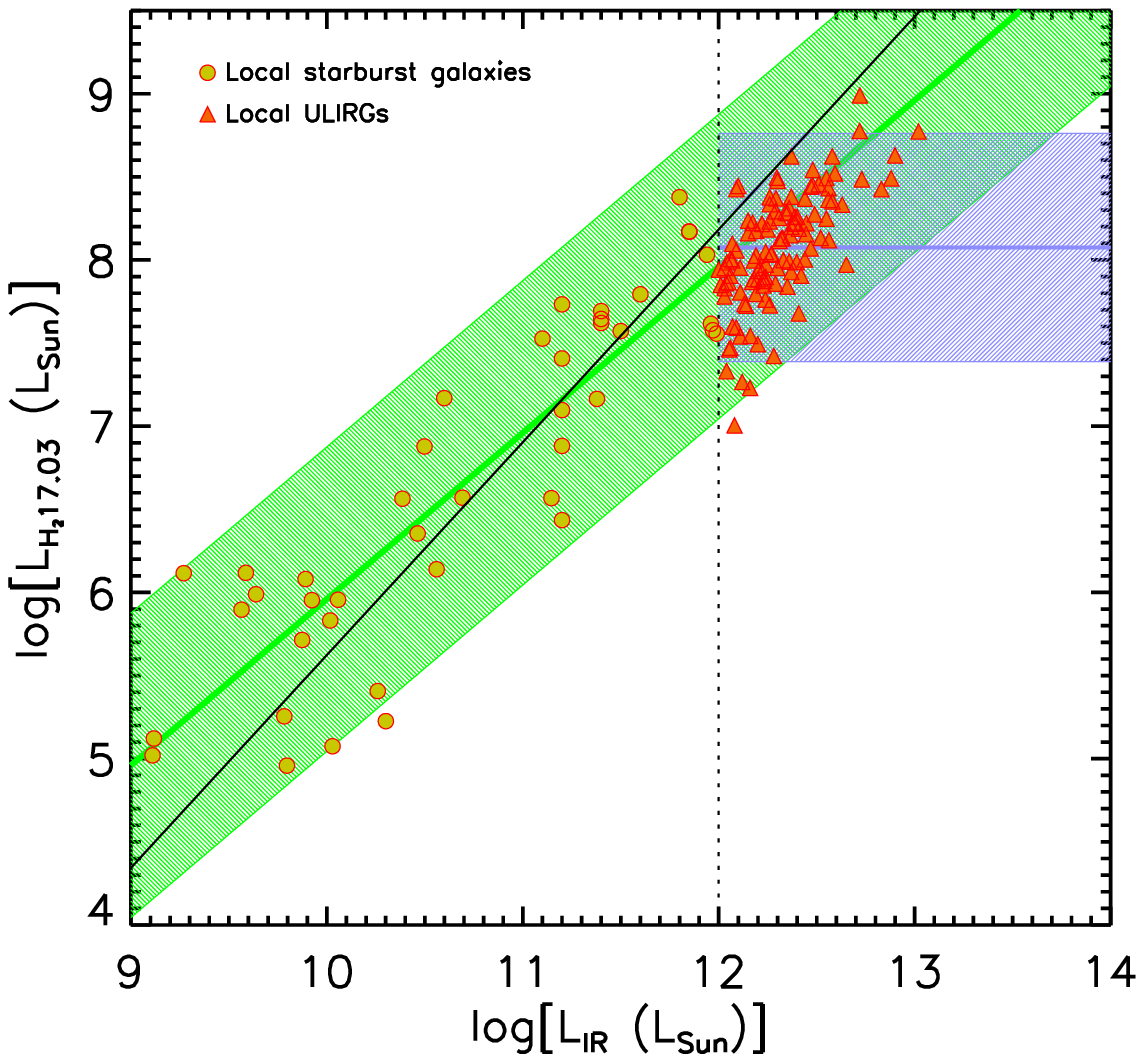}
\includegraphics[trim=3.45cm 1.1cm 1.4cm 1.2cm,clip=true,width=0.45\textwidth]{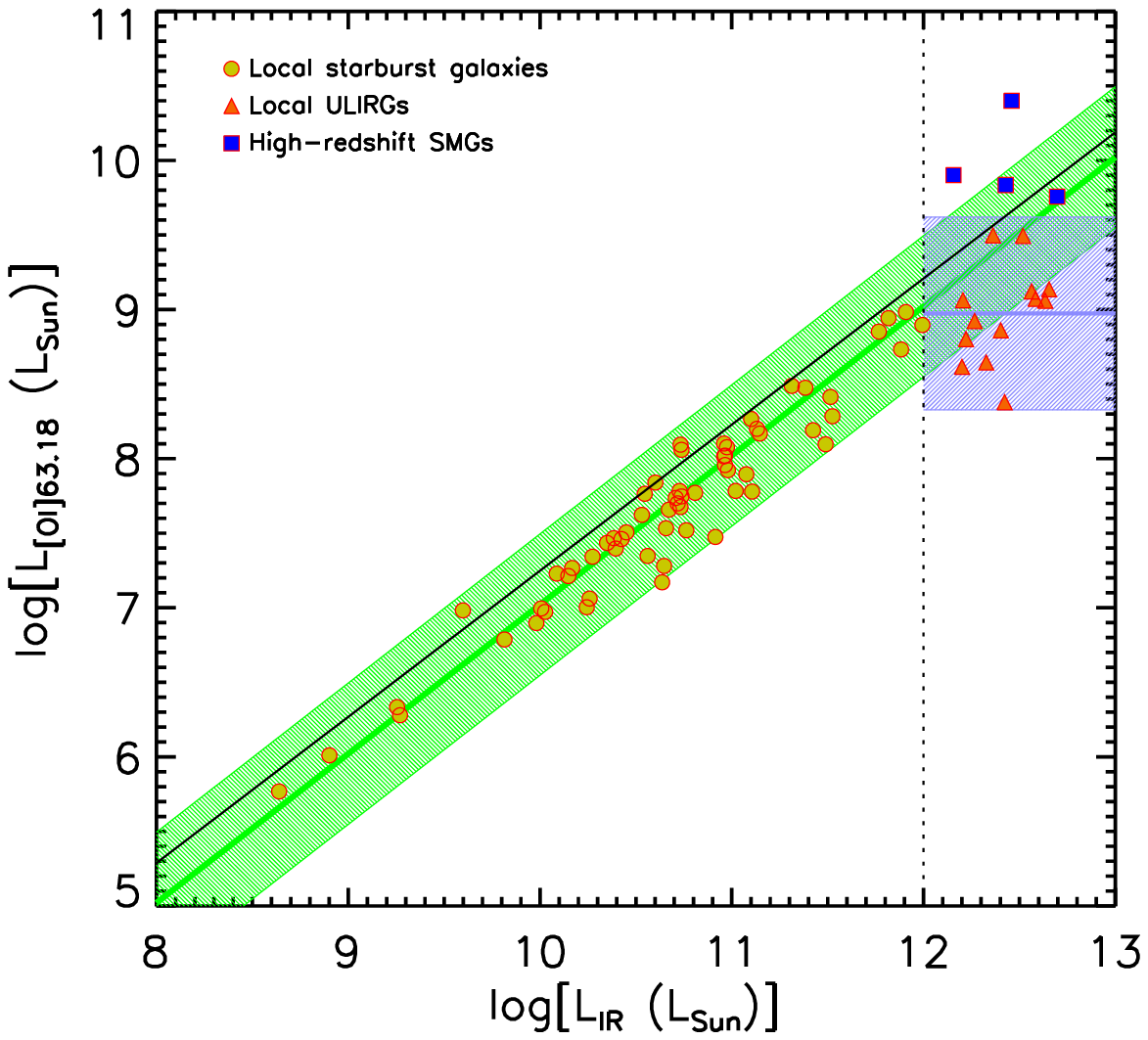}
\includegraphics[trim=3.45cm 1.1cm 1.4cm 1.2cm,clip=true,width=0.45\textwidth]{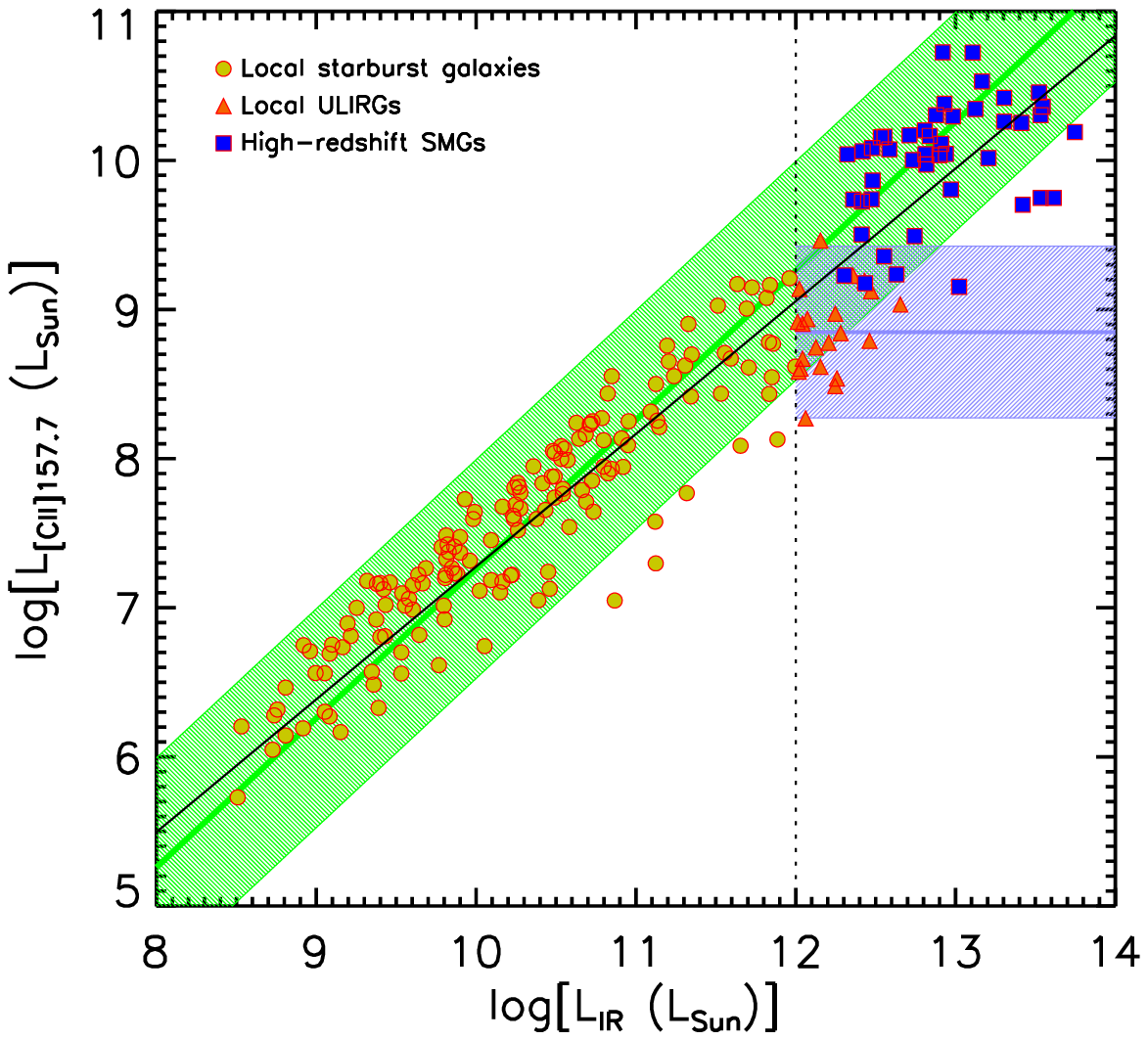}
%
%
\caption{Luminosity of the PAH\,$11.25\,\mu$m (top-left panel), H$_{2}$\,17.03$\mu$m (top-right panel), [OI]\,$63.18\mu$m (bottom-left panel) and [CII]\,$157.7\mu$m (bottom-right panel) lines, versus continuum IR luminosity. The \emph{green band} shows the $2\,\sigma$ range around the mean linear relation $\log(L_{\ell})=\log(L_{\rm IR}) + c$ for local star forming galaxies with $L_{\rm IR}<10^{12}\,L_{\odot}$ (\emph{circles}) and high-redshift SMGs (\emph{squares}); the values of $c\equiv\langle \log(L_{\ell}/L_{\rm IR})\rangle$ are given in Table\,\ref{tab:c_d_values}. The \emph{black lines} show the best-fit relations derived by \citet{Spin12}. The \emph{azure band} shows the $2\,\sigma$ spread around the mean line luminosity for the sample of local ULIRGs (\emph{triangles}) whose line luminosities appear to be uncorrelated with $L_{\rm IR}$ and are generally lower than expected from the linear relation holding for the other sources.  The mean line luminosities $\langle \log L_{\ell}\rangle$ of these objects are given in Table\,\ref{tab:c_d_values}. The \emph{vertical dotted black line} at $L_{\rm IR}=10^{12}\,L_{\odot}$ marks the lower boundary of ULIRG luminosities \citep{SandersMirabel1996}.}
 \label{fig:new_cal}
  \end{center}
\end{figure*}

\begin{figure*}
\includegraphics[trim=0.15cm 2.0cm 0.6cm 2.3cm,clip=true,width=\textwidth]{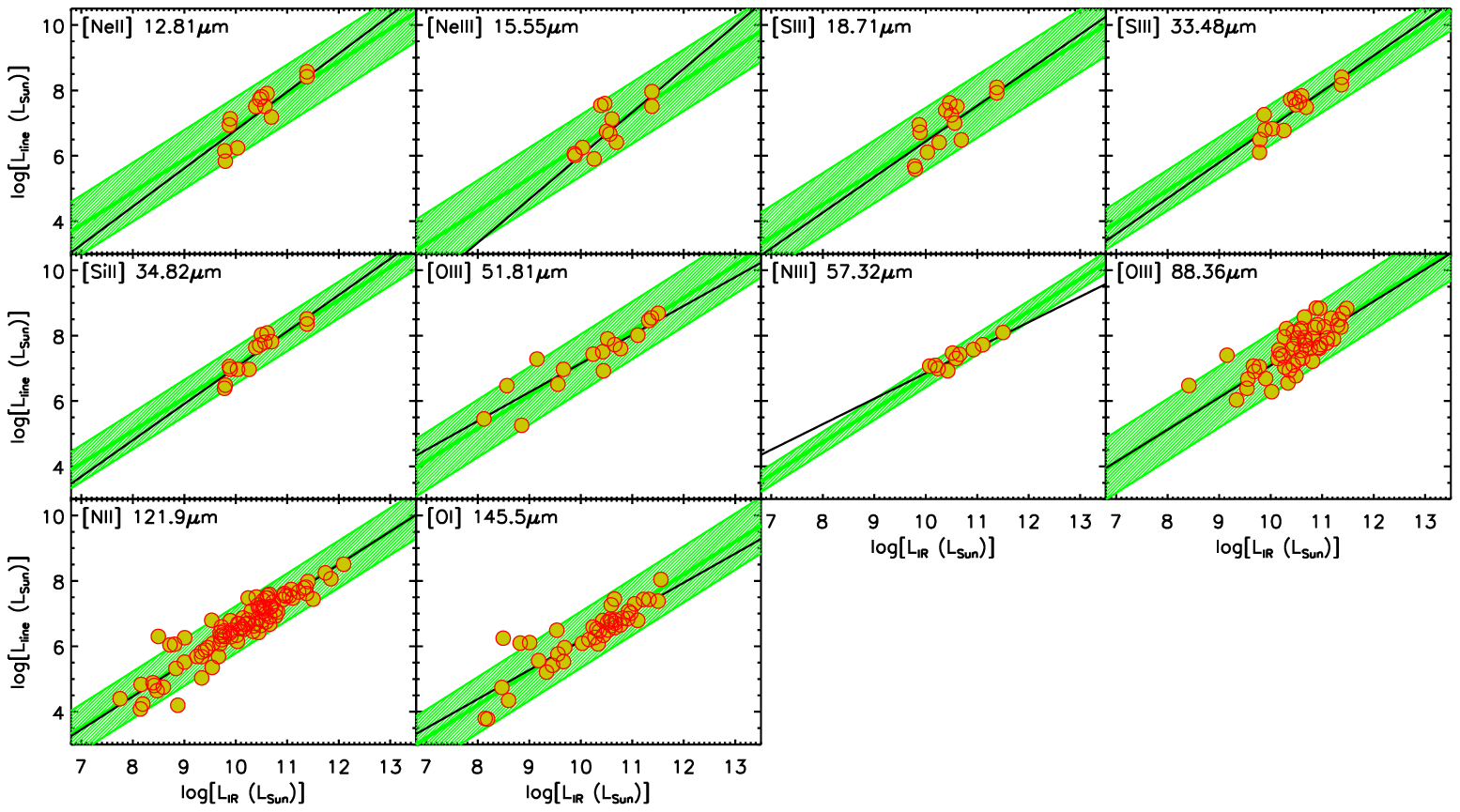} 
\caption{Log of the line luminosity as a function of $\log(L_{\rm IR})$, in solar units, for all the IR lines studied in the present work, except for the four lines (PAH\,$11.25\,\mu$m, H$_{2}$\,17.03\,$\mu$m, [OI]\,$63.18\,\mu$m and [CII]\,$157.7\,\mu$m) considered separately (see Fig.~\protect\ref{fig:new_cal}). Data for $\lambda \le 34.82\,\mu$m refer to the local starburst galaxies from the \citet{Bern09} catalogue. Data at longer wavelengths are from the heterogeneous sample of \citet{Brauher08}. The \emph{green band} shows the $2\,\sigma$ spread around the mean linear relation $\log(L_{\ell})=\log(L_{\rm IR}) + c$ (represented by the \emph{green line}). The \emph{black lines} show the best-fit relations derived by \citet{Spin12}. Our simulations (see Sect.~\protect\ref{sect:simul}) favour a direct proportionality between line and continuum luminosity. }
 \label{fig:line_vs_IR}
\end{figure*}

\begin{table}
\centering
\footnotesize
\begin{tabular}{lcc}
\hline
\hline
\rule[-3mm]{0mm}{6mm}
Spectral line & $\displaystyle\big\langle\log\big(\displaystyle{L_{\ell}\over L_{\rm IR}}\big)\big\rangle\ (\sigma)$ & $\langle \log(L_{\ell})_{\rm UL}\rangle\ (\sigma)$ \\
\hline
${\rm PAH 11.25}\mu$m       &       -2.29\ (0.36)       &      9.01\ (0.28)   \\
${\rm [NeII]}12.81\mu$m      &       -3.11\ (0.45)       &    -   \\
${\rm [NeIII]}15.55\mu$m     &       -3.69\ (0.47)       &     -   \\
${\rm H_{2}}17.03\mu$m      &      -4.04\ (0.46)       &      8.07\ (0.34)   \\
${\rm [SIII]}18.71\mu$m       &        -3.49\ (0.48)       &     -   \\
${\rm [SIII]}33.48\mu$m       &        -3.05\ (0.31)       &     -   \\
${\rm [SiII]}34.82\mu$m       &         -2.91\ (0.28)       &     -   \\
${\rm [OIII]}51.81\mu$m      &        -2.84\ (0.44)       &     -   \\
${\rm [NIII]}57.32\mu$m      &       -3.26\ (0.16)       &     -   \\
${\rm [OI]}63.18\mu$m        &          -2.99\ (0.24)       &      8.97\ (0.32)   \\
${\rm [OIII]}88.36\mu$m      &       -2.87\ (0.47)       &     -   \\
${\rm [NII]}121.9\mu$m       &        -3.49\ (0.36)       &     -   \\
${\rm [OI]}145.5\mu$m        &         -3.80\ (0.43)       &     -   \\
${\rm [CII]}157.7\mu$m      &          -2.74\ (0.37)       &      8.85\ (0.29)   \\
\hline
\hline
\end{tabular}
\caption{Mean values of the log of line to IR (8-1000\,$\mu$m) continuum luminosities, $\langle\log({L_{\ell}/L_{\rm IR}})\rangle$, and associated dispersions $\sigma$. For the PAH$\,11.25\,\mu$m, H$_{2}\, 17.03\,\mu$m, ${\rm [OI]}\,63.18\,\mu$m and ${\rm [CII]}\,157.7\,\mu$m  lines, $\langle\log({L_{\ell}/L_{\rm IR}})\rangle$ has been computed excluding local ULIRGs, for which the luminosity in these lines appears to be uncorrelated with $L_{\rm IR}$. For the latter objects the last column gives the mean values of $\log(L_{\ell})$ in these lines and their dispersions (the luminosities are in solar units).}
\label{tab:c_d_values}
\end{table}

%

\begin{figure*}
\makebox[\textwidth][c]{
\includegraphics[trim=2.4cm 1.2cm 2.7cm 0.3cm,clip=true,width=\textwidth]{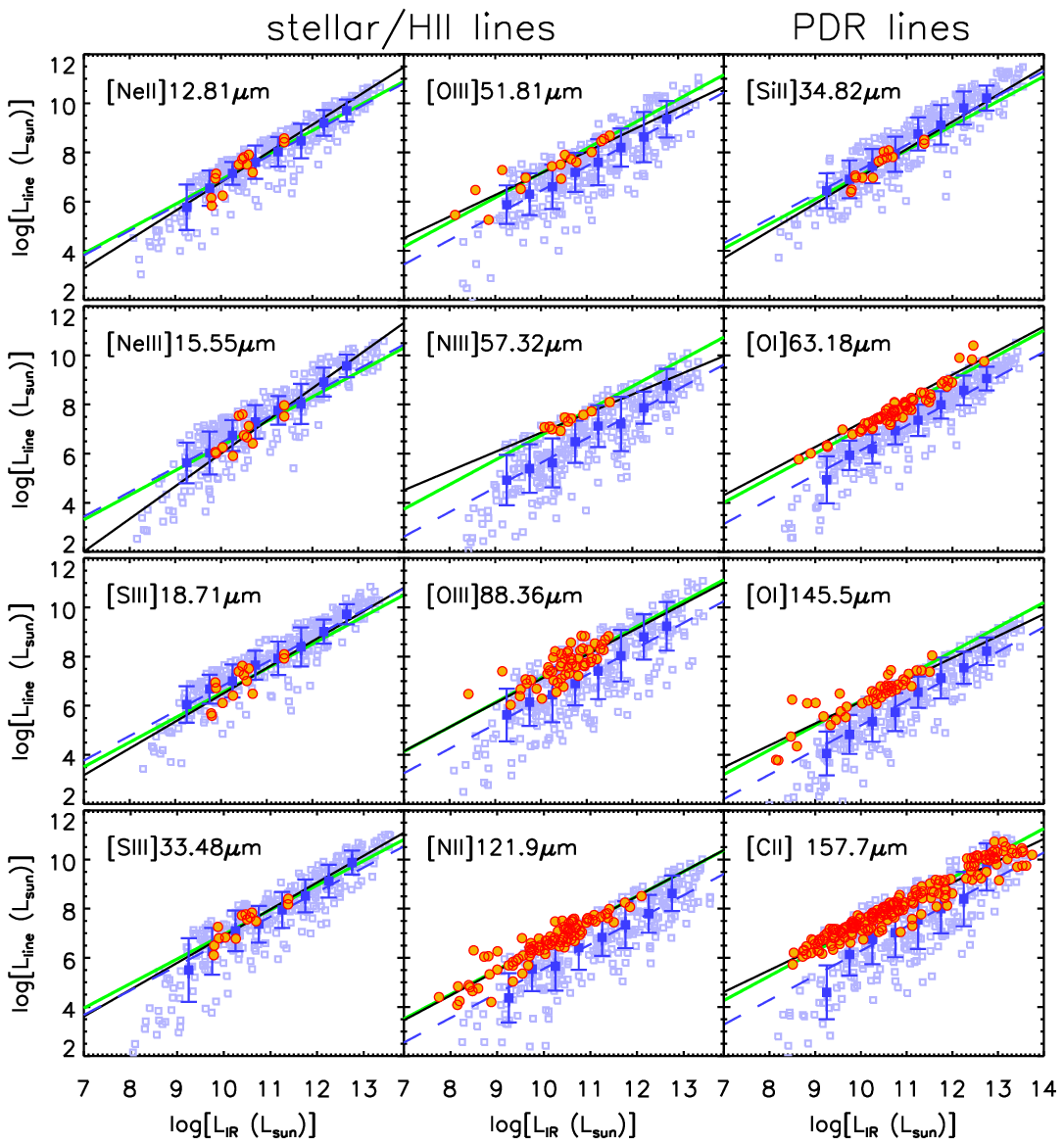}}
\caption{Simulated line versus IR luminosity (\emph{open squares}) for some fine structure lines produced in stellar/H{\small II} regions (left and central panels) and in PDRs (right panels). Only purely star forming galaxies (no AGN contributions) are considered in the simulations.
The \emph{dashed blue lines} correspond to the average ratios $c=\langle \log(L_{\ell}/L_{\rm IR})\rangle$, i.e. to a direct proportionality between $\log(L_{\ell})$ and $\log(L_{\rm IR})$, for the simulated data points after binning them in steps of 0.25 in $\log L_{\rm dust}$ (\emph{filled squares} with error bars). The \emph{solid green lines} correspond to the average ratios for the real data points, the \emph{solid black lines} show the best-fit relations derived by \citet{Spin12}. The data points (local star forming galaxies and high-z SMGs) are the same as in Figs.~\ref{fig:new_cal} and \ref{fig:line_vs_IR}. The simulations are systematically low for longer wavelength lines ($\lambda > 50\,\mu$m) as expected since they do not include the emission from outside H{\small II} regions. 
}
 \label{fig:Lline_vs_Lir_simul}
\end{figure*}
%

\section{Properties of emitting regions from infrared lines}\label{sect:IRlines}


Emission line intensities and emission line ratios in the mid- and far-IR domain, that do not suffer as much from dust extinction as optical and UV emission lines, supply unique information on the physical conditions (electron density and temperature, degree of ionization and excitation, chemical composition) of the line emitting gas in the dominant dust-obscured regions of galaxies with intense star formation activity \citep{Spin92,Rubin94,Panuzzo2003}. For example, different IR fine structure transitions of the same ion have different sensitivity to collisional de-excitation. This can be used to identify the typical electron density of the emitting gas. A typical line pair used for this purpose is [OIII]$52/88\,\mu$m. Furthermore, the relative strengths of the fine-structure emission lines in different ionization stages of suitable elements may be used to probe the ionizing spectral
energy distributions (SEDs) because different ionization stages are formed by significantly different photon energy ranges in the ionizing spectrum. Thus these line ratios are sensitive to the hardness of the stellar energy distribution, hence to the
most massive stars present \citep[e.g.][]{Thornley2000}. Therefore they provide information on the stellar initial mass function (IMF), on the age of the ionizing stellar population and on the ionization parameter in single HII regions or in starbursts of short duration. Also measuring these ratios can constrain the emission from non-stellar sources of ionizing photons, such as active galactic nuclei (AGNs). A typical ratio used for this purpose is [NeIII]$15.55\mu$m/[NeII]$12.81\mu$m. 

Some of the FIR lines are important cooling lines, like the [CII]157.7$\mu$m, that is the primary coolant of the warm, neutral ISM, and the [OI]63.18$\mu$m and [OIII]88.36$\mu$m lines that become more important than [CII]157.7$\mu$m at high densities and temperatures.

The Polycyclic Aromatic Hydrocarbons (PAH) mid-IR emission lines are presumably originated in very large carbon-rich ring molecules or in very small amorphous carbon dust grains \citep{Puget89}. Analyzing galactic IR spectra associated to different physical stages (photo-dissociation regions, HII regions, planetary nebulae and quiescent spirals), \citet{Galliano06} found that PAH line ratios are a powerful diagnostic tool of the physical conditions inside the region where the emission is originating. The PAH content is remarkably correlated with the rate of carbon dust production by AGB stars, consistent with the notion that PAHs form in
the envelopes of these stars. AGB stars have a lifetime of several hundred million years and inject their material into the ISM when the system is already evolved. Thus PAH lines are a sort of chronometer constraining the age of the stellar populations.

$\hbox{H}_{2}$, the most abundant molecule in the Universe, is very useful to study the interstellar medium and the star formation processes. In fact the formation of $\hbox{H}_{2}$ on grains starts the chemistry of interstellar gas and then this molecule gives an important contribution to the cooling of astrophysical media, in particular in low-metallicity and low-temperature environments.

Besides the PAH\,11.25\,$\mu$m and H$_{2}$\,17.03\,$\mu$m lines, the set of IR fine-structure emission lines studied in this paper includes 4 lines from photodissociation regions ([SiII]\-34.82\,$\mu$m, [OI]\-63.18\,$\mu$m, [OI]\-145.5\,$\mu$m and [CII]\-157.7\,$\mu$m) and 8 stellar/HII region lines ([NeII]\-12.81\,$\mu$m, [NeIII]\-15.55\,$\mu$m, [SIII]\-18.71\,$\mu$m, [SIII]\-33.48\,$\mu$m, [OIII]\-51.81\,$\mu$m, [NIII]\-57.32\,$\mu$m, [OIII]\-88.36\,$\mu$m and [NII]\-121.9\,$\mu$m).

Finally, we note that line profiles can be used to study the dynamics of the ionized gas.

\section{Line versus IR luminosity}\label{sect:line_vs_IR}

We have updated the relations between line and IR luminosities in several respects. First we have found and collected from the literature additional measurements for the  PAH\,$11.25\,\mu$m, H$_{2}\,17.03\,\mu$m, ${\rm [OI]}\,63.18\,\mu$m and ${\rm [CII]}\,157.7\,\mu$m lines.

For the PAH\,$11.25\,\mu$m line we have used data on: local star forming galaxies from \citet{Bern09,Brandl06,Brandl09,O'Dowd09,O'Dowd11} and \citet{Pereira-Santaella10}; local ULIRGs from \citet{Imanishi07,Imanishi09,Imanishi10}; high-$z$ SMGs from \citet{Sajina07,Yan05,Yan07,Pope08,Fiolet10}.

For the H$_{2}\,17.03\,\mu$m line, data on: local star forming galaxies from  \citet{Devost04,Roussel06,Farrah07,Bern09,Brandl09,Veilleux09,Pereira-Santaella10,Cormier12}; local ULIRGs from \citet{Higdon06,Farrah07,Veilleux09}.

For the [OI]\,$63.18\,\mu$m line, data on: local star forming galaxies from \citet{Colbert99,Malhotra01,Negishi01,Gracia11,Cormier12}; local ULIRGs from \citet{Luhman2003,Fischer2010,Gracia11}; high-$z$ SMGs from \citet{Coppin12}.

Finally for the [CII]\,$157.7\,\mu$m line, data on: local star forming galaxies from \citet{Carral94,Colbert99,Unger00,Malhotra01,Negishi01,Gracia11,Cormier12} and \citet{Swinbank12}; local ULIRGs from \citet{Luhman98,Luhman2003,Gracia11} and \citet{Swinbank12}; high-$z$ SMGs from \citet{Colbert99,Maiolino2009,Hailey-Dunsheath10,Ivison10,Stacey10,Wagg10,Cox11,DeBreuck11,Gracia11,Swinbank12,Walter12,Riechers13}  and \citet{CarilliWalter2013}.

From these samples we have excluded all objects for which there is evidence for a substantial AGN effect on the strength of the lines.

When far-IR luminosities over rest-frame wavelength ranges different from the one (8--1000\,$\mu$m) adopted here were given we applied the following conversions, from \citet{Stacey10} and \citet{GraciaCarpio2008}, respectively
\begin{eqnarray}
L_{\rm FIR}(40-500\,\micron)&=&1.5 \times L_{\rm FIR}(42-122\,\micron),\\
L_{\rm IR}(8-1000\,\micron)&=&1.3  \times L_{\rm FIR}(40-500\,\micron).
\end{eqnarray}
The correlations between line luminosities and $L_{\rm IR}$ are shown in Fig.\,\ref{fig:new_cal}.

A common property of these four lines is that they are not (or not only) produced in HII regions but in (or also in) neutral and ionized interstellar medium and in photo-dissociation regions (PDRs). In fact, as pointed out by \citet{Panuzzo2003}, because carbon has an ionization potential (11.26 eV) lower than that of H, the CII ion is present in PDRs and in the neutral medium illuminated by far-UV stellar radiation. Indeed, the [CII]\,$157.7\,\mu$m line is the most important coolant of the warm neutral medium. The excitation temperature of the [OI]\,$63.2\,\mu$m line is of $228\,$K \citep{Kaufman1999} so that it can be easily produced in the neutral medium. The H$_{2}\,17.03\,\mu$m line is the strongest molecular hydrogen line used to give information on the conditions of the warm component in the PDRs. Polycyclic aromatic hydrocarbon (PAH) molecules are found to be ubiquitous in the interstellar medium (ISM) but apparently do not survive in the ionized gas \citep{Tielens2008}. An important consequence is that these lines suffer from much less extinction than those produced inside the dense, dust enshrouded stellar birth clouds.

Moreover, these lines are less directly linked to the SFR (and therefore to $L_{\rm IR}$) than lines originated deep in stellar birth clouds. This difference may have to do with the steep drop in the $L_{\rm [CII]}/L_{\rm IR}$ ratio observed in local ULIRGs  \citep{Luhman2003,Gracia11} while measurements of this ratio in high-redshift galaxies of similar luminosity is in the range observed for local galaxies with $L_{\rm IR}\lsim 10^{11}\,L_\odot$ \citep{Maiolino2009,Stacey10,George2013}, although this finding may result from a selection effect. The estimated duration of the star formation episode in local ULIRGs, generally triggered by interactions/mergers, is $\sim 0.1\,$Gyr, i.e. of the same order as the typical time \citep[as estimated by][]{Silva98} for hot stars to leave the dense birth clouds, where UV photons capable of ionizing C atoms have a short mean free path for absorption. Thus PDRs are necessarily small and $L_{\rm [CII]}$ correspondingly low. When the birth clouds begin to dissolve and larger PDRs can be produced the SFR is declining, in many cases below the threshold for the classification as ULIRGs.

In massive high-$z$ galaxies the star formation is galaxy-wide and, as shown by \citet{Lapi11}, the data on luminosity functions indicate a longer duration ($\sim 0.5\hbox{--}0.7\,$Gyr) of the intense star formation phase. Thus a significant fraction of hot stars have the time to leave the birth clouds and to migrate to less dense regions where they can generate extended PDRs while the SFR is still very high. This can explain their larger $L_{\rm [CII]}/L_{\rm IR}$ ratios. The same argument holds for non-ULIRG low-$z$ galaxies whose star formation is long lived. Alternative explanations are discussed by \citet{CarilliWalter2013}.

Anyway, the luminosity of low-$z$ ULIRGs in these lines appears to be essentially uncorrelated with $L_{\rm IR}$. For these objects we have adopted Gaussian distributions of the logarithm of the line luminosity, $\log(L_{\ell})$, around a mean value, $\langle \log(L_{\ell})\rangle$, independent of $L_{\rm IR}$. On the contrary, the mean line luminosities for the other low- and high-$z$ galaxies, and for all galaxies in the cases of the other lines considered in this paper, are found to be tightly correlated with, and essentially proportional to $L_{\rm IR}$ (Figs.\,\ref{fig:new_cal} and \ref{fig:line_vs_IR}). Again we have adopted Gaussian distributions around $\langle \log(L_{\ell}/L_{\rm IR})\rangle$. The different relationships between line and continuum luminosities for low-$z$ ULIRGs have however only a marginal impact on the predicted number of line detections. If we apply also to these galaxies the relationships found for the other populations the results change by less than 2\%.

\subsection{Simulations}\label{sect:simul}

The data on line and continuum luminosities refer to heterogeneous samples of galaxies, affected by unknown selection effects and unknown and variable amounts of dust extinction. They may therefore not be representative of the true distributions of line-to-IR luminosity ratios. In other words, the available measurements do not provide an unambiguous determination of the line-IR luminosity relationship. Also in some cases the statistics is plainly insufficient. To get some guidance on the choice of the slope of the relationship in the light of the relevant astrophysics we have carried out simulations taking into account the expected emission line intensities for different properties (density, metallicity, filling factor) of the emitting gas, different ages of the stellar populations, and the range of dust obscuration. 

We have used the public library\footnote{http://pasquale.panuzzo.free.fr/hii/} of line luminosities compiled by \citet{Panuzzo2003}. This library provides luminosities  for 60 nebular emission lines, at wavelengths from the UV to the far-IR, and for 54 H and He recombination lines. There are 12 lines in common with those considered by \citet{Spin12}. 
Eight of them are generated in star formation/H{\small II} regions while the others may be associated with PDRs. The library was produced interfacing version 94 of the photo-ionization code CLOUDY \citep{Ferland01} with the single stellar population (SSP) model of \citet{Bressan94} which provides the spectrum of the ionizing sources. The analysis by \citet{Panuzzo2003} showed that the emission line spectrum of an H{\small II} region with fixed gas properties  (metallicity, density and geometry) is described with reasonable precision by only three quantities: the production rates of photons capable of ionizing H{\small I}, He{\small I} and O{\small II}:
\begin{eqnarray}
Q_{\rm H} = \int_{\nu_{\rm H}}^{\infty}\frac{L_{\nu}}{h\nu}d\nu, ~
Q_{\rm He} = \int_{\nu_{\rm H}e}^{\infty}\frac{L_{\nu}}{h\nu}d\nu, ~
Q_{\rm O} = \int_{\nu_{\rm O}}^{\infty}\frac{L_{\nu}}{h\nu}d\nu,
\end{eqnarray}
where $L_{\nu}$ is monochromatic luminosity of the SSP (in units of $\hbox{erg}\,\hbox{s}^{-1}\,\hbox{Hz}^{-1}$) and $\lambda_{\rm H} = c/\nu_{\rm H} = 911.76$\,\AA, $\lambda_{\rm He} = c/\nu_{\rm He} = 504.1$\,{\AA} and $\lambda_{\rm O} = c/\nu_{\rm O} = 350.7$\,{\AA} are the photo-ionization threshold wavelengths for H{\small I}, He{\small I} and O{\small II}, respectively. Therefore, different ionizing sources that provide the same values of $Q_{\rm H}$, $Q_{\rm He}$ and $Q_{\rm O}$ produce the same emission line spectra, with reasonable accuracy.

\begin{figure}
\hspace{-4.4cm}
\makebox[\textwidth][c]{
\includegraphics[trim=0.05cm 0.0cm 0.0cm 0.0cm,clip=true,width=0.8\textwidth, angle=0]{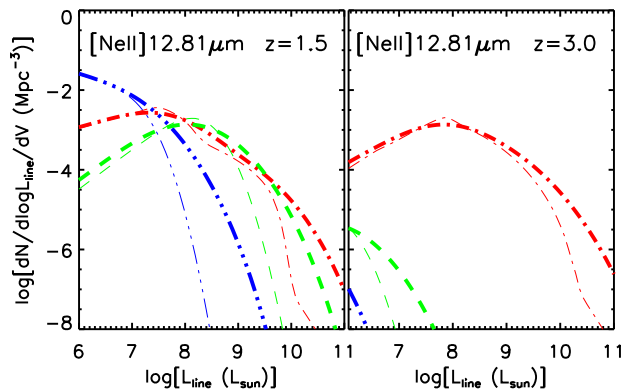}
}
\vspace{-3.7cm}
\caption{Effect of the dispersion in the line-to-IR luminosity ratio on the predicted luminosity function of the {\rm [NeII]}12.81$\mu$m line. \emph{Thick lines}: including dispersion; \emph{thin lines}: no dispersion. Contributions of the different source populations: \emph{dash-dot-dotted blue lines} for cold late-type galaxies, \emph{green dashed lines} for warm late-type galaxies and \emph{dash-dotted red lines} for proto-spheroids.
}
\label{fig:disp_NeII12.81}
\end{figure}

The \citet{Panuzzo2003} library provides line luminosities for a grid of values of $Q_{\rm  H}$, $Q_{\rm He}$ and $Q_{\rm O}$, and a range of values of the gas density, $\rho_{\rm gas}$, inside the star birth clouds, the gas metallicity, $Z_{\rm gas}$, and the gas
filling factor, $\epsilon_{\rm gas}$. According to the authors the range of parameters covered by their library is fully adequate to describe the emission properties of the majority of star forming galaxies. They do not consider line emission from PDRs
and from diffuse warm neutral/low ionized medium, although some fine structure IR lines are efficiently produced in these media. As a
consequence their luminosities for these lines should be taken as lower limits.
\begin{figure*}
\hspace{+0.0cm}
\makebox[\textwidth][c]{
\includegraphics[trim=0.45cm 0.8cm 0.9cm 1.3cm,clip=true,width=1.1\textwidth, angle=0]{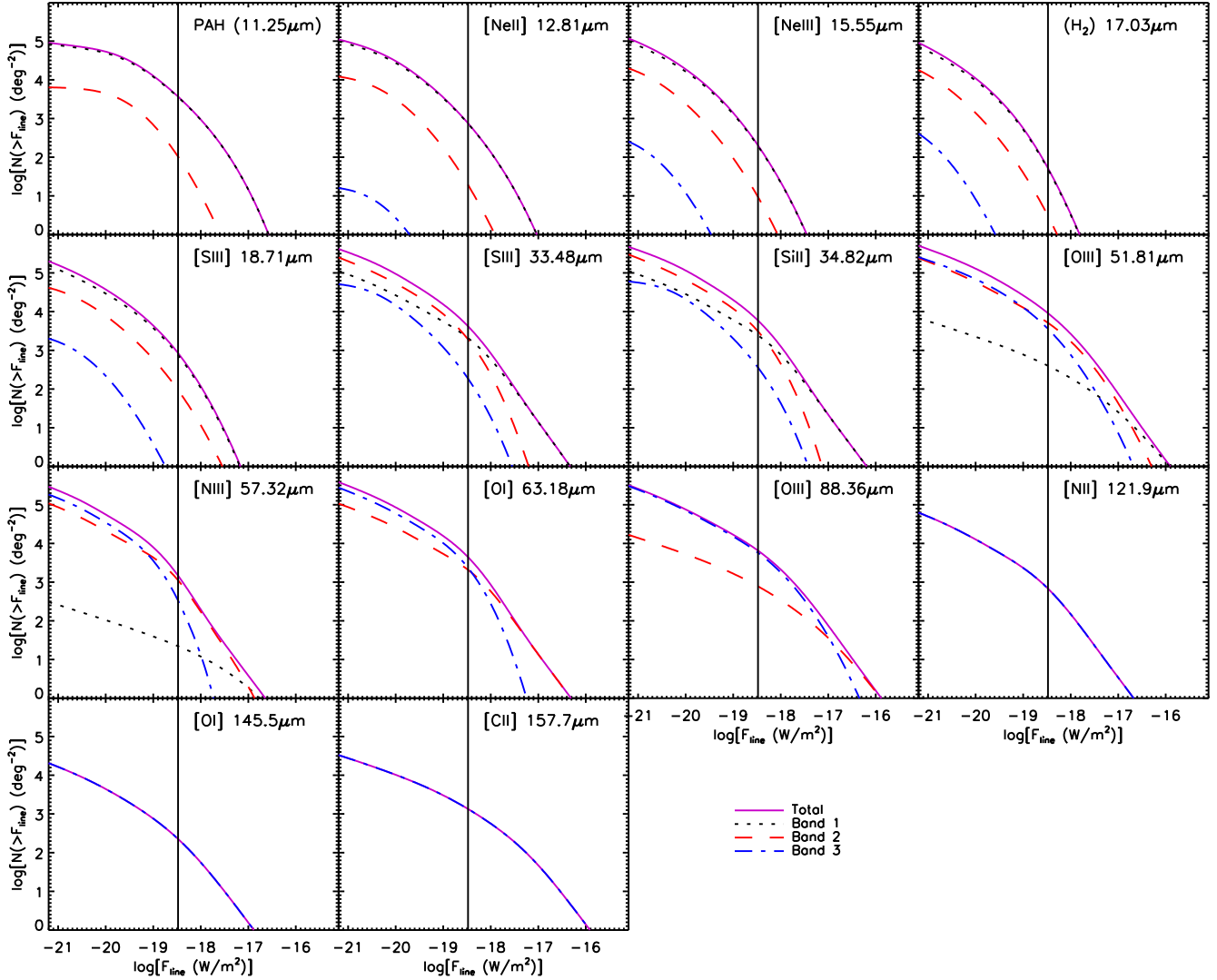}
}
\caption{Integral counts of star forming galaxies as a function of line fluxes over the full SPICA-SAFARI wavelength range (\emph{violet solid line}) and in each of its three bands. The \emph{vertical lines} correspond to the average detection limit in the 3 bands for the reference survey.}
 \label{fig:intcounts_spica_all}
\end{figure*}

To estimate the distribution of the line-to-IR luminosity ratios we simulated the SED of star forming galaxies in the absence of dust extinction and derived both the line luminosities,  using the \citet{Panuzzo2003} library, and the continuum IR luminosity, assuming an extinction law. The SED was generated using GALAXEV, the library of evolutionary stellar population synthesis of \cite{BC03}. The continuum emission at the time $t_{\rm obs}$ of a stellar population characterized by a star formation rate SFR$(t)$, assuming that the dust absorption only depends on the stellar age, is written as \citep[e.g.][]{CF00}:
\begin{equation}
L_{\lambda}^{\rm SED}(t_{\rm obs}) = \int_{0}^{t_{\rm obs}} {\rm SFR}(t_{\rm
  obs}-t) L_{\lambda}^{\rm SSP}(t)T_{\lambda}(t)dt,
\end{equation}
where $L_{\lambda}^{\rm SSP}(t)$ is the power radiated per unit interval of frequency and of initial mass by the SSP of age $t$, $T_{\lambda}(t)$ is the ``transmission function'' defined as the fraction of the radiation, produced at wavelength $\lambda$ and at the time $t$, that escapes from the galaxy. The fraction absorbed by dust in the galaxy is therefore $1-T_{\lambda}(t)$. The transmission function is written as $T_{\lambda}(t) = \exp[-\tau_{\lambda}(t)]$, $\tau_{\lambda}(t)$ being the effective absorption optical depth at wavelength $\lambda$ seen by stars at age $t$. Stars are born in dense molecular clouds, which dissipate typically on a time-scale $t_{\rm BC}$. This implies that the emission from stars younger than $t_{\rm BC}$ is more attenuated than that from older stars. \citet{CF00} write the transmission function as
\begin{equation}
T_{\lambda}(t) = \left\{
\begin{array}{l}
T_{\lambda}^{\rm BC}  \times T_{\lambda}^{\rm ISM}  ~~~ \mbox{ for $t\leq t_{\rm BC}$}, \\
~~ \\
T_{\lambda}^{\rm ISM}  ~~~~~~~~~~~~ \mbox{ for $t>t_{\rm BC}$} \\
 \end{array} \right.
\end{equation}
where $T_{\lambda}^{\rm BC}$ and $T_{\lambda}^{\rm ISM}$ are the fractions of radiation transmitted by the birth clouds and by the
ambient inter-stellar medium (ISM), respectively, here assumed (for simplicity) to be independent of time. We model the effective optical depths of the birth clouds and of the ISM, $\tau_{\lambda}^{\rm BC}$ and $\tau_{\lambda}^{\rm ISM}$, respectively, as in \citet{daCunha08}:
\begin{equation}\label{eq:tauV_BC}
\tau_{\lambda}^{\rm BC} = (1-\mu)\tau_{V} \left( \frac{\lambda}{5500{\rm
      \AA}} \right)^{-1.3}
\end{equation}
\begin{equation}
\tau_{\lambda}^{\rm ISM} = \mu\tau_{V} \left( \frac{\lambda}{5500{\rm \AA}} \right)^{-0.7},
\end{equation}
$\tau_{V}$ being the total effective $V$-band absorption optical depth of the dust seen by young stars inside the birth clouds and
$\mu=\tau_{V}^{\rm ISM}/(\tau_{V}^{\rm BC} + \tau_{V}^{\rm ISM})$ the fraction of this contributed by dust in the ambient ISM.

The total luminosity absorbed by dust in the birth clouds and reradiated at far-IR/sub-millimeter wavelengths is \citep{daCunha08}
\begin{eqnarray}
L_{\rm dust}^{\rm BC}(t_{\rm obs}) = \int_{0}^{\infty}\!\!\!\!\!\!\!\! {\rm d}\lambda (1 - T_{\lambda}^{\rm BC})
\int_{0}^{t_{\rm BC}}\!\!\!\!\!\!\!\! {\rm d}t\,\, {\rm SFR}(t_{\rm
  obs}-t) L_{\lambda}^{\rm SSP}(t), \nonumber
\end{eqnarray}
\begin{equation}
~
\end{equation}
while the total luminosity absorbed by the {\it ambient ISM} is
\begin{eqnarray}
L_{\rm dust}^{\rm ISM}(t_{\rm obs}) = \int_{0}^{\infty} \!\!\!\!\!\!\!\! {\rm d}\lambda (1 - T_{\lambda}^{\rm ISM})
\int_{t_{\rm BC}}^{t_{\rm obs}} \!\!\!\!\!\!\!\! {\rm d}t\, {\rm SFR}(t_{\rm
  obs}-t) L_{\lambda}^{\rm SSP}(t). \nonumber
\end{eqnarray}
\begin{equation}
~
\end{equation}
Therefore the total luminosity absorbed and re-radiated by dust is
\begin{equation}
L_{\rm dust}(t_{\rm obs}) = L_{\rm dust}^{\rm BC}(t_{\rm
  obs}) + L_{\rm dust}^{\rm ISM}(t_{\rm obs}).
\end{equation}
We take this as the total infrared luminosity $L_{\rm IR}$.

The luminosity of a nebular line of wavelength $\lambda_{\ell}$ is given by
\begin{equation}\label{eq:L_line}
L_{\ell}(t_{\rm obs}) = \int_{0}^{t_{\rm BC}} {\rm SFR}(t_{\rm obs}-t) L_{\ell}^{\rm SSP}(t)T_{\lambda_{\ell}}(t)\,{\rm d}t,
\end{equation}
where $L_{\ell}^{\rm SSP}(t)$ is the luminosity of the line produced in the H{\small II} regions by a stellar generation of age $t$.
GALAXEV was used to compute the SSP continuum luminosity, $L_{\lambda}^{\rm SSP}(t)$, from which the $Q$-values are derived as a function of the SSP age and used to get the line luminosities from Panuzzo's library. Within GALAXEV we adopt the SSP model computed with the evolutionary tracks and isochrones by \citet{Bertelli1994}, with solar metallicity (i.e. $Z=Z_{\odot}=0.02$) and a \citet{Chabrier2003} IMF. We assume an exponentially declining star formation rate, $\hbox{SFR}(t)= (A_{\rm SFR}/\tau)\, \exp(-t/\tau)$, with $\tau=10^{7}\,$yr (=$t_{\rm BC}$). The normalization, $A_{\rm SFR}$, corresponds to the stellar mass assembled by the time $t>>\tau$. Line luminosities were simulated for different values of $t_{\rm
  obs}$, $A_{\rm SFR}$, $\rho_{\rm gas}$, $\epsilon_{\rm gas}$, $\tau_{V}$ and $\mu$
within the range specify below, assuming uniform distributions,
\begin{itemize}
\item $\log (t_{\rm obs}/{\rm yr})\in \,[5.0-8.0]$,
\item $\log(A_{\rm SFR}/M_{\odot})\,\in\,[7.0-11.0]$,
\item $\rho_{\rm gas}\in \,[10-10000]$ and $\epsilon_{\rm gas}\in\,[0.001-1.0]$, as these are the intervals considered by \citet{Panuzzo2003},
\item  $\log (\tau_{V})\in\,[\log 2-2]$ as in \cite{daCunha10},
\item $\mu\tau_{V}\in\,[0-2]$ as in \cite{daCunha10}.
\end{itemize}
%
%

We made 300 simulations for each line. The derived line luminosities are shown in Fig.\,\ref{fig:Lline_vs_Lir_simul} as a function of the continuum IR luminosity and compared with the available data.
The solid green lines and the dashed blue lines correspond to a direct proportionality between $\log(L_{\ell})$ and $\log(L_{\rm IR})$ i.e. to the average ratios $c=\langle \log(L_{\ell}/L_{\rm IR})\rangle$ for real and simulated data, respectively. For simulated data the ratios were computed  after binning them in steps of 0.25 in $\log L_{\rm dust}$.

Overall, we find a good agreement between observed and simulated distributions of luminosities for lines with $\lambda < 50\,\mu$m. For these lines the simulations are consistent with a direct proportionality between $L_{\ell}$ and $L_{\rm IR}$ and the derived values of $c=\langle \log(L_{\ell}/L_{\rm IR})\rangle$ are consistent with the determinations directly based on the data. On the other hand, the luminosities of longer wavelength lines found from simulations fall systematically below the observed luminosities as expected, since, as mentioned above, the \citet{Panuzzo2003} library refers to the emission line spectrum of H{\small II} regions only.

Based on these results, we have adopted the mean $\log(L_{\ell}/L_{\rm IR})$, or the mean $\log(L_{\ell})$ in the case of low-$z$ ULIRGs, and the associated dispersions derived from observations, buying from simulations the indication of a direct proportionality between line and IR luminosity. As illustrated by Fig.\,\ref{fig:Lline_vs_Lir_simul} the relationships obtained in this way differ from those derived by \citet{Spin12} from direct fits of the data. Differences are small whenever the data points are numerous and span a broad range of $L_{\rm IR}$ but may be substantial in other cases (up to a factor of 10 at the highest and lowest luminosities). This illustrates the benefit of exploiting astrophysical inferences, as per our simulations.

\begin{figure}
\hspace{-3.5cm}
\makebox[\textwidth][c]{
\includegraphics[trim=3.1cm 0.5cm 1.2cm 3.2cm,clip=true,width=0.75\textwidth, angle=0]{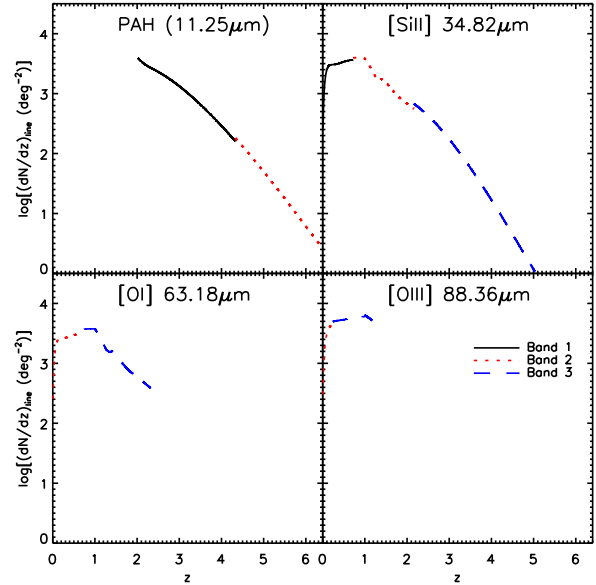}
}
\vspace{-3.0cm}
\caption{Examples of the predicted redshift distributions of sources detected by SPICA/SAFARI for a 1\,hr exposure per FoV. The different colours correspond to the 3 spectral bands. }
 \label{fig:redshiftdistr_spica}
\end{figure}
%

\begin{figure}
\hspace{-3.5cm}
\makebox[\textwidth][c]{
\includegraphics[trim=3.1cm 0.95cm 1.2cm 3.2cm,clip=true,width=0.75\textwidth, angle=0]{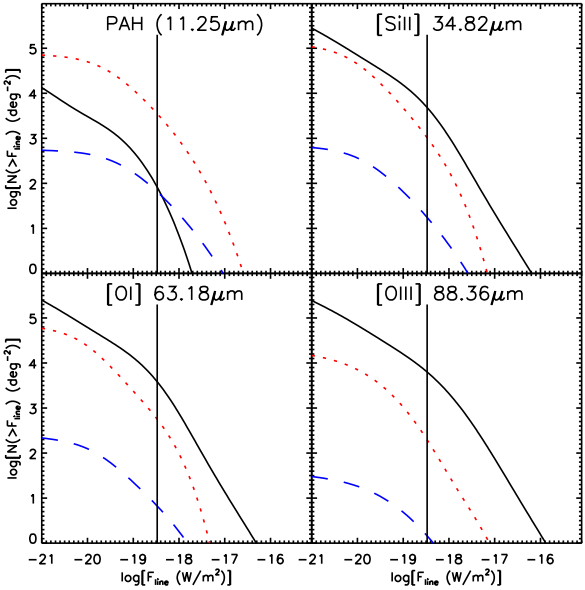}
}
\vspace{-2.5cm}
\caption{Contributions of different galaxy populations (\emph{solid black line}: late-type -- starburst and spiral -- galaxies; \emph{dotted red line}: unlensed proto-spheroids; \emph{dashed blue line}: gravitationally lensed proto-spheroids) to the integral counts as a function of fluxes in the PAH\,$11.25\micron$, [SiII]\,$34.82\micron$, [OI]$\,63.18\micron$ and [OIII]$\,88.36\micron$ lines, within the SPICA/SAFARI wavelength range. The \emph{vertical lines} correspond to the average value of the detection limits ($3.7\times10^{-19}\,\hbox{W}/\hbox{m}^{2}$, $3.4\times10^{-19}\,\hbox{W}/\hbox{m}^{2}$, $2.9\times10^{-19}\,\hbox{W}/\hbox{m}^{2}$) for  the three bands ($\,34-60\,\mu$m, $\,60-110\,\mu$m, $\,110-210\,\mu$m) for 1-hr exposure per FoV.}
 \label{fig:intcounts_spica}
\end{figure}

\section{Line luminosity functions and number counts}\label{sect:LF}

The line luminosity functions can be easily computed from the IR ones by convolving the latter with the distribution of the IR-to-line luminosity ratios, $p[\log(L_{\rm IR}/L_{\ell})]$, i.e.
\begin{equation}
\frac{{\rm d}\mathcal{N}^{3}(L_{\ell},z)}{{\rm d}\log{L_{\ell}}\, {\rm d}z \, {\rm d}V} = \int_{-\infty}^{+\infty}  \Phi(L_{\ell},z) p[\log(L_{\rm IR}/L_{\ell})] \log{L_{\rm IR}},
\end{equation}
where $\Phi(L_{\rm IR},z)$ is the comoving luminosity function per unit ${\rm d}\log L_{\rm IR}$, and $p$ is assumed to be a Gaussian with mean value and dispersion given in Table~\ref{tab:c_d_values}. However we do not use this formalism but adopt instead a Monte Carlo approach in which simulated catalogues are constructed from the theoretical IR luminosity functions. This approach has the advantage of providing information on the luminosity (and flux) of {\it all} the lines, at the same time, for each simulated object and therefore allows us to predict the number of galaxies detectable in two or more lines in future spectroscopic surveys.

The simulation is performed as follows. For a given galaxy population and a given redshift bin, $\Delta z$, we draw from the corresponding IR luminosity function a number of IR luminosities equal to that of the objects expected over a (reference) area $\Omega=0.5$\,deg$^{2}$, and with IR luminosity greater than a minimum value that we set\footnote{This value is chosen to ensure that all the galaxies detectable by SAFARI and CCAT are represented in the simulation while keeping the computing time, that depends on the total number of simulated sources, to an acceptable level.} to $\log(L_{\rm IR, min}/L_{\odot})=7.0$, i.e.
\begin{equation}
\label{eq:flux_function}
\mathcal{N}(z,\Omega) = \int_{\log L_{\rm IR, min}}^{+\infty} \!\!\!\!\!\!\Phi(L_{\ell},z)\,{\rm d}\log L_{\rm IR}\, \frac{{\rm d}^2V_{c}(z)}{{\rm d}z\,{\rm d}\Omega}\, \Delta z\, \Omega,
\end{equation}
where ${\rm d}V_{c}(z)$ is the comoving volume element.

We then associate to each simulated IR source a line luminosity by sampling, at random, the distribution of values of $\log(L_{\ell}/L_{\rm IR})$ or of $\log(L_{\ell})_{\rm UL}$, assumed to be Gaussian with mean value and dispersion given in Table~\ref{tab:c_d_values}. Similarly we associate a redshift to each object, assuming a uniform probability distribution within $z\pm(\Delta z/2)$. Finally the line flux, $F_{\ell}$, is computed from the simulated redshift, $z_{\rm simul}$, and the simulated line luminosity, $L_{\ell,\rm simul}$, as $F_{\ell,\rm simul}=L_{\ell,\rm simul}/4\pi d^{2}_{\rm L}(z_{\rm simul})$, $d_{L}(z)$ being the luminosity distance.
The wavelength range covered by the instrument, $[\lambda_{\rm min}, \lambda_{\rm max}]$, and the line rest-frame wavelength, $\lambda_{\ell}$, define the minimum and maximum redshift within which the line is detectable, $z_{\ell, \rm min/max} = \lambda_{\rm min/max}/\lambda_{\ell} - 1$. When $z_{\rm simul}$ falls outside the range defined by $z_{\ell, \rm min}$ and $z_{\ell, \rm max}$, the corresponding line luminosity and line flux are set to 0.

The simulated line luminosities are binned to produce the line luminosity functions at a given redshift. In order to reduce the effect of fluctuations in the highest luminosity bins, where the statistic is poorest, the whole procedure is repeated 300 times and the simulated luminosity functions are averaged together.
The predicted number counts and redshift distributions are also derived by averaging over the 300 simulated catalogues.

\begin{figure*}
\makebox[\textwidth][c]{
\includegraphics[trim=2.3cm 0.8cm 0.7cm 0.2cm,clip=true,width=0.48\textwidth, angle=0]{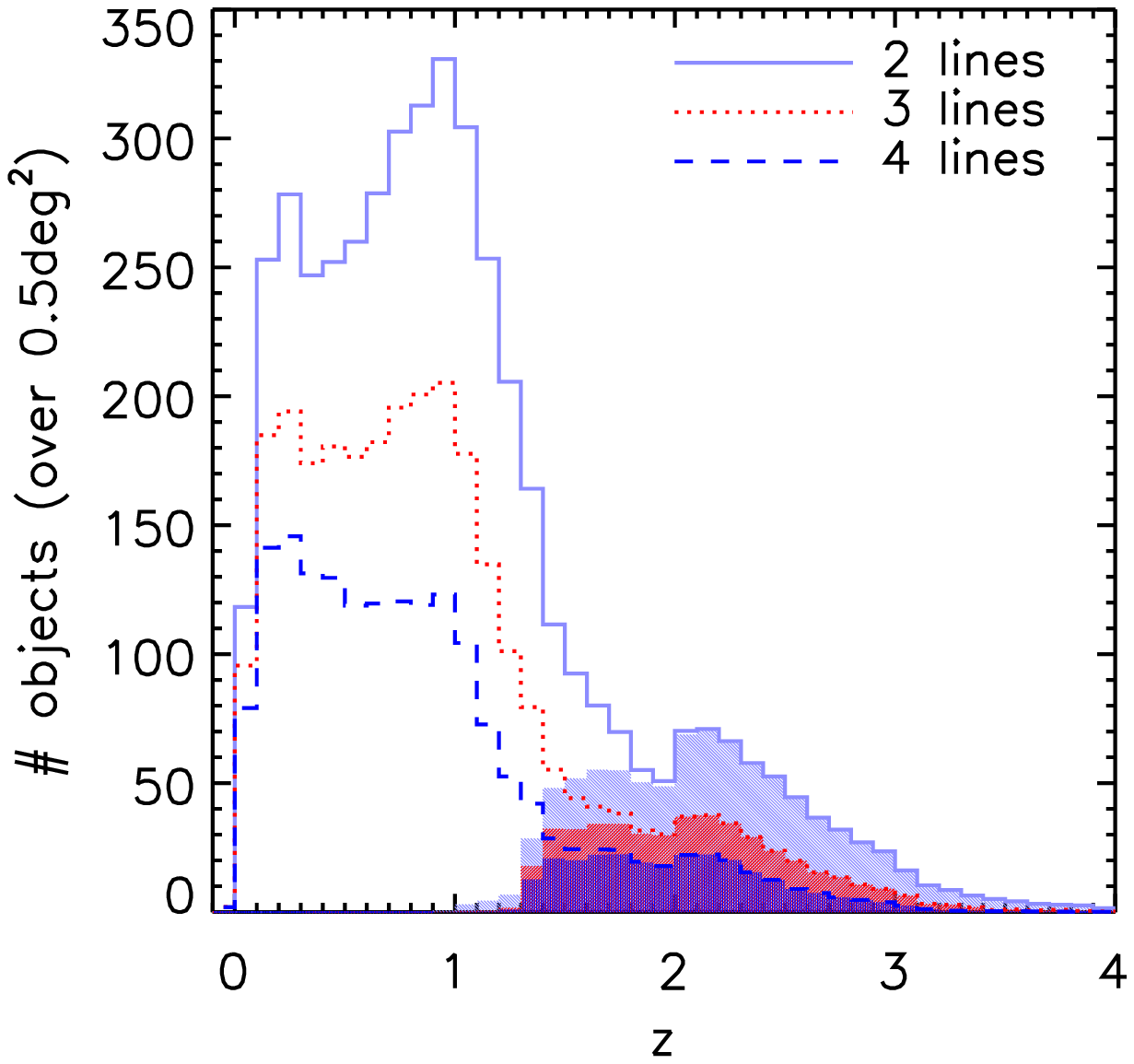}
\includegraphics[trim=2.3cm 0.8cm 0.7cm 0.2cm,clip=true,width=0.48\textwidth, angle=0]{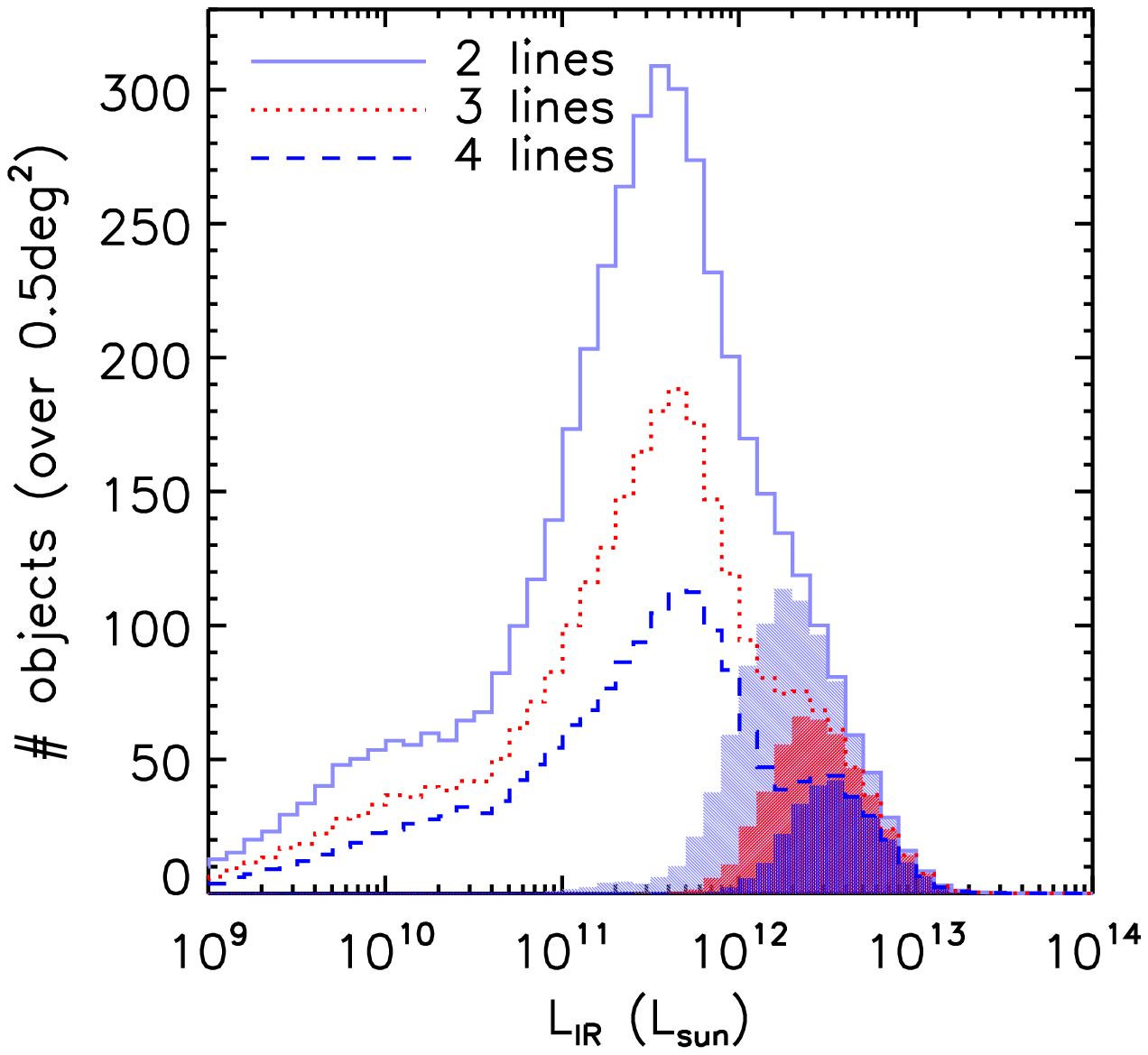}
}
\caption{Predicted redshift (left) and luminosity (right) distributions of star forming galaxies detectable in at least two (\emph{cyan solid histogram}), three (\emph{red dotted}) and four (\emph{blue dashed}) spectral lines, by the reference survey with SPICA/SAFARI. The \emph{shaded areas} represent the contributions of proto-spheroids.}
 \label{fig:z_distr_tot}
\end{figure*}
%

%
\begin{table*}
\centering
\footnotesize
\begin{tabular}{lcccccccc}
\hline
\hline
Spectral line & $0.00-0.75$ & $0.75-1.25$ & $1.25-1.75$ & $1.75-2.25$
& $2.25-2.75$ & $2.75-4.00$ & $4.00-6.00$ & All z \\
\hline
{\rm PAH11.25}$\mu$m &                0 &                0 &                0 &              378 &              597 &              548 &               83 &            1606 \\
{\rm [NeII]12.81}$\mu$m &                0 &                0 &               47 &               146 &               75 &               49 &                4 &        321 \\
{\rm [NeIII]15.55}$\mu$m &                0 &                8 &               46 &                21 &                9 &                5 &                0 &                89 \\
{\rm H$_{2}$17.03}$\mu$m &                0 &                9 &                10 &                4 &                2 &                0 &                0 &               25 \\
{\rm [SIII]18.71}$\mu$m &                0 &              195 &               108 &               48 &               26 &               12 &                1 &             390 \\
{\rm [SIII]33.48}$\mu$m &              919 &              661 &              285 &              125 &              78 &               41 &                2 &               2111 \\
{\rm [SiII]34.82}$\mu$m &             1222 &             957 &              436 &              191 &              117 &              64 &                2 &       2989 \\
{\rm [OIII]51.81}$\mu$m &             1735 &             1490 &             1071 &              423 &              230 &               78 &                0 &               5027 \\
{\rm [NIII]57.32}$\mu$m &              498 &              218 &              91 &               51 &               22 &                0 &                0 &            880 \\
{\rm [OI]63.18}$\mu$m &             1082 &             879 &              378 &              163 &               17 &                0 &                0 &              2519 \\
{\rm [OIII]88.36}$\mu$m &             1907 &             1611 &              325 &                0 &                0 &                0 &                0 &           3843 \\
{\rm [NII]121.9}$\mu$m &              415 &                0 &                0 &                0 &                0 &                0 &                0 &            415 \\
{\rm [OI]145.5}$\mu$m &               140 &                0 &                0 &                0 &                0 &                0 &                0 &              140 \\
{\rm [CII]157.7}$\mu$m &              807 &                0 &                0 &                0 &                0 &                0 &                0 &        807 \\
\hline
\hline
\end{tabular}
\caption{Predicted redshift distributions of star forming galaxies detectable by the SPICA/SAFARI reference survey ($0.5\,\hbox{deg}^2$, 1\,hr integration per FoV) in each of the 14 emission lines considered in this paper.}
\label{tab:counts_spica1}
\end{table*}

\begin{table*}
\centering
\footnotesize
\begin{tabular}{lcccc}
\hline
\hline
\small{Spectral line} & \textbf{\small{\citet{Cai13}}} & \small{\citet{Franc10}} & \small{\citet{Grupp11}} & \small{\citet{Valiante09}} \\
\hline
{\rm PAH11.25}$\mu$m & \textbf{1606} & 160 & 173 & 418 \\
{\rm [NeII]12.81}$\mu$m & \textbf{321} & 333 & 203 & 42 \\
{\rm [NeIII]15.55}$\mu$m & \textbf{89} & 129 & 122 & 179 \\
{\rm H$_{2}$17.03}$\mu$m & \textbf{25} & 13 & 17 & 6 \\
{\rm [SIII]18.71}$\mu$m & \textbf{390} & 74 & 67 & 6 \\
{\rm [SIII]33.48}$\mu$m & \textbf{2111} & 1377 & 1059 & 1103 \\
{\rm [SiII]34.82}$\mu$m & \textbf{2989} & 2535 & 2241 & 3037 \\
{\rm [OIII]51.81}$\mu$m & \textbf{5027} & 1723 & 2081 & 3883 \\
{\rm [NIII]57.32}$\mu$m & \textbf{880} &  276 & 665 & 973 \\
{\rm [OI]63.18}$\mu$m & \textbf{2519} & 4171 & 5421 & 8897 \\
{\rm [OIII]88.36}$\mu$m & \textbf{3843} & 2286 & 3027 & 5121 \\
\hline
\hline
\end{tabular}
\caption{Comparison of the number of $5\,\sigma$ detections in each line by SPICA/SAFARI according to the present model (second column, in boldface) with those for the 3 models used by \citet{Spin12}. The predictions for the \citet{Franc10} model ($3^{rd}$ column) refer (as ours) to  star forming galaxies only, while those for the \citet[][$4^{th}$ column]{Grupp11} and for the \citet[][$5^{th}$ column]{Valiante09} model include also AGNs.}
\label{tab:counts_spica_tot}
\end{table*}

\begin{table*}
\centering
\footnotesize
\begin{tabular}{lcccccccc}
\hline
\hline
Spectral line & $0.75-1.25$ & $1.25-1.75$ & $1.75-2.25$
& $2.25-2.75$ & $2.75-4.00$ & $4.00-6.00$ & $6.00-8.00$ & All z \\
\hline
{\rm [OIII]51.81}$\mu$m &             0 &             0 &              0 &              0 &               0 &                14 &     2 &          16 (1) \\
{\rm [OI]63.18}$\mu$m &                 0 &              0 &          0 &               0 &                2 &                12 &      0 &        14 (0) \\
{\rm [OIII]88.36}$\mu$m &                 0 &              0 &          0 &             328 &               701 &               197 &   2 &       1228 (0) \\
{\rm [NII]121.9}$\mu$m &                 0 &                137 &       172 &            126 &           75 &               11 &     6 &       527 (0) \\
{\rm [OI]145.5}$\mu$m &               61 &                160 &        140 &           43 &              15 &                47 &     2 &      468 (2) \\
{\rm [CII]157.7}$\mu$m &                3379 &                3099 &     2865 &          50 &                679 &                2243 &    0 &    12315 (12) \\
\hline
\hline
\end{tabular}
\caption{Predicted redshift distributions of star forming galaxies detectable by CCAT over $1.0\,\hbox{deg}^2$ with 1\,hr integration per FoV in each of the emission lines within the CCAT wavelength range. In parenthesis we give the number of detections of strongly lensed galaxies at $z>5$. No detections in the [NIII]\,57.32\,$\mu$m line are expected at the flux limits of this survey.}
\label{tab:counts_ccat}
\end{table*}

\begin{table*}
\centering
\footnotesize
\begin{tabular}{lcccccccc}
\hline
\hline
\small{Spectral line} & \textbf{\small{\citet{Cai13}}} & \small{\citet{Franc10}} & \small{\citet{Grupp11}} & \small{\citet{Valiante09}} \\
\hline
{\rm [OI]63.18}$\mu$m & \textbf{7 (1)} & 0 & 2 & 0 \\
{\rm [OIII]88.36}$\mu$m & \textbf{614 (515)} & 133 & 182 & 278 \\
{\rm [NII]121.9}$\mu$m & \textbf{264 (255)} & 29 & 54 & 92 \\
{\rm [OI]145.5}$\mu$m & \textbf{234 (210)} & 1 & 5 & 2 \\
{\rm [CII]157.7}$\mu$m & \textbf{6158 (5036)} & 4421 & 4683 & 6612 \\
\hline
\hline
\end{tabular}
\caption{Numbers of $5\,\sigma$ detections by CCAT for the survey considered by \citet{Spin12} (1\,h integration per FoV over an area of $0.5\,\hbox{deg}^2$) as predicted by the present model (second column, in boldface) compared with those given (only for 5 lines) by \citet{Spin12} for the 3 models used by them. The latter predictions include the contributions of AGNs and are limited to $z\le 4$. For a fairer comparison we also give, in parenthesis, our predictions for the same redshift range.}
\label{tab:counts_ccat_tot}
\end{table*}

\begin{table*}
\centering
\footnotesize
\begin{tabular}{lccccccccc}
\hline
\hline
Spectral line & $t=6\,\hbox{min}$ & $t=30\,\hbox{min}$ & \boldmath$t=1\,\hbox{h}$ & $t=2\,\hbox{h}$ & $t=3\,\hbox{h}$ & $t=5\,\hbox{h}$ & $t=10\,\hbox{h}$ \\
  & $5.00\,\hbox{deg}^2$ & $1.00\,\hbox{deg}^2$ & \boldmath$0.50\,\hbox{deg}^2$ & $0.25\,\hbox{deg}^2$ & $0.17\,\hbox{deg}^2$ & $0.10\,\hbox{deg}^2$ & $0.05\,\hbox{deg}^2$ \\
\hline
{\rm PAH11.25}$\mu$m & 3721 (709)   & 2124 (525) &  \textbf{1606 (441)} &         1180 (356) &                977 (307) &                765 (251) &              536 (184) \\
{\rm [NeII]12.81}$\mu$m & 500 (31) & 389 (34) &   \textbf{321 (31)} &          262 (31) &                231 (30) &            191 (26) &               145 (22) \\
{\rm [NeIII]15.55}$\mu$m & 82 (0) & 98 (2) &   \textbf{89 (2)} &         81 (3) &                76 (3) &               70 (3) &                59 (3) \\
{\rm H$_{2}$17.03}$\mu$m & 19 (0) & 19 (0) &  \textbf{25 (0)} &           24 (1) &                23 (0) &                24 (1) &                22 (1) \\
{\rm [SIII]18.71}$\mu$m & 497 (9)  & 453 (8) &    \textbf{390 (9)} &        334 (9) &              297 (10) &               248 (8) &               191 (7) \\
{\rm [SIII]33.48}$\mu$m & 3773 (11) & 2634 (27) &  \textbf{2111 (24)} &     1617 (28) &              1355 (25) &              1066 (24) &              748 (22) \\
{\rm [SiII]34.82}$\mu$m & 5862 (30) & 3863 (36) &  \textbf{2989 (39)} &       2232 (39) &             1842 (38) &              1422 (35) &              978 (31) \\
{\rm [OIII]51.81}$\mu$m & 14003 (12) & 7054 (14) &  \textbf{5027 (12)} &       3488 (9) &             2785 (7) &             2072 (6) &              1378 (5) \\
{\rm [NIII]57.32}$\mu$m & 971 (0) & 953 (0) &    \textbf{880 (0)} &      754 (0) &              678 (0) &              572 (0) &               434 (0) \\
{\rm [OI]63.18}$\mu$m & 4376 (0) & 3181 (0) &   \textbf{2519 (0)} &     1891 (0) &             1568 (0) &              1209 (0) &              826 (0) \\
{\rm [OIII]88.36}$\mu$m & 12391 (0) & 5661 (0) &   \textbf{3843 (0)} &       2554 (0) &             1996 (0) &              1452 (0) &                930 (0) \\
{\rm [NII]121.9}$\mu$m & 886 (0) & 548 (0) &    \textbf{415 (0)} &      308 (0) &                252 (0) &                195 (0) &                132 (0) \\
{\rm [OI]145.5}$\mu$m & 322 (0) & 187 (0) &    \textbf{140 (0)} &       102 (0) &                83 (0) &                64 (0) &                44 (0) \\
{\rm [CII]157.7}$\mu$m & 3276 (0) & 1247 (0) &   \textbf{807 (0)} &       515 (0) &                397 (0) &                280 (0) &                175 (0) \\
\hline
\hline
\end{tabular}
\caption{Number of star forming galaxies detectable by a SPICA/SAFARI survey of 450\,hours covering different areas, with correspondingly different integration times ($t$) per FoV. In parenthesis we give the number of detections at $z>3$.}
\label{tab:counts_spica2}
\end{table*}

In Fig.~\ref{fig:disp_NeII12.81} we show, as an example, the effect of the dispersion on the predicted luminosity function for the [NeII]12.81$\mu$m line. Since dispersions are substantial, they result in a vast increase, compared to the case without dispersion, of the density of the most luminous sources and, correspondingly, of the bright portion of source counts.

The integral counts predicted by our model in the SPICA/SAFARI wavelength range, for all the 14 lines considered in this work, are shown in Fig.~\ref{fig:intcounts_spica_all}.

According to B. Sibthorpe (private communication), the SPICA/SAFARI reference $5\sigma$ detection limits for an integration of 1\,hr per FoV are $3.7\times10^{-19}\,\hbox{W}/\hbox{m}^{2}$ for the first band ($\,34-60\,\mu$m), $3.4\times10^{-19}\,\hbox{W}/\hbox{m}^{2}$ for the second band ($\,60-110\,\mu$m) and $2.9\times10^{-19}\,\hbox{W}/\hbox{m}^{2}$ for the third band ($\,110-210\,\mu$m).  The predicted numbers of sources brighter than these limits in each line, detected by the survey considered by \citet{Spin12}, covering $0.5\,\hbox{deg}^2$ (hereafter `the reference survey') are given in the last column of Table~\ref{tab:counts_spica1}. The other columns detail the redshift distributions of detected sources. Examples of such redshift distributions are displayed, for 4 lines, in Fig.~\ref{fig:redshiftdistr_spica}, where the redshift ranges covered by the different bands are identified by different colours. Obviously, only the shortest wavelength lines can be detected up to very high redshifts. According to our model, the reference survey  can detect the PAH$\,11.25\,\mu$m line in some galaxies at redshifts up to $\simeq 6$. With the exception of 3 lines ([NeIII]\,15.55\,$\mu$m, H$_{2}$\,17.03\,$\mu$m, [SIII]\,18.71\,$\mu$m) out of 14 the slope of the SPICA/SAFARI integral counts, just below the mean of the detection limits of the three bands, of the reference survey is $< 2$ so that the number of detections for given observing time increases more with the surveyed area than with survey depth. And also for these three lines, the slope flattens below 2 not far from these limit.

Figure~\ref{fig:intcounts_spica} shows the contributions to the counts for the 4 lines in Fig.~\ref{fig:redshiftdistr_spica} of late-type (spiral plus starburst) galaxies and of proto-spheroids, lensed and unlensed. At the $5\sigma$ SPICA/SAFARI limits for 1\,hr exposure we expect the detection of the  PAH$\,11.25\,\mu$m of $\sim 60$ strongly lensed proto-spheroids per square degree.
Figure~\ref{fig:z_distr_tot} illustrates the luminosity and redshift distributions of sources for which the SPICA/SAFARI reference survey will detect at least 2, 3 or 4 lines. From our model we expect about 4600 sources detected by the SPICA/SAFARI reference survey in at least 2 lines, about 2800 in at least 3 lines and about 1800 in at least 4 lines. The numbers of proto-spheroids among these are about 900, 460 and 270, respectively.

\begin{figure*}
\hspace{+0.0cm}
\includegraphics[trim=0.7cm 1.0cm 1.3cm 1.2cm,clip=true,width=0.99\textwidth, angle=0]{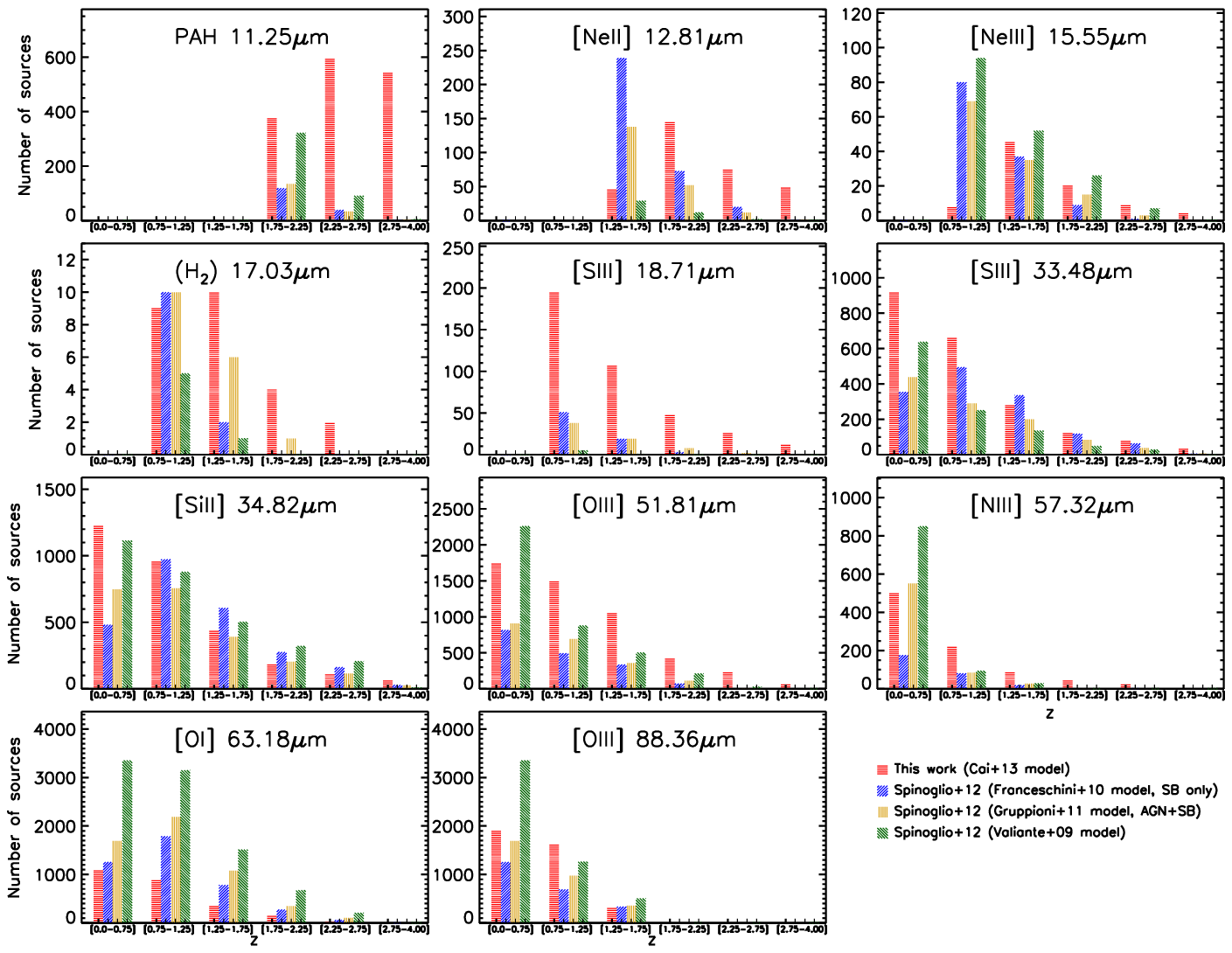}
\vspace{+0.0cm}
\caption{Redshift distributions of galaxies detectable in 11 lines by the SPICA/SAFARI reference survey. The predictions of our model (\emph{bars with red horizontal dashes}) are compared with those of the 3 models used by \citet{Spin12}; see the legend on the bottom right.}
 \label{fig:histoSPICA}
\end{figure*}

Figure~\ref{fig:histoSPICA} compares the present predictions for the redshift distributions of sources detected by the SPICA/SAFARI reference survey in 11 lines with those from the 3 models used by \citet{Spin12}.
The main differences concern the PAH\,11.25\,$\mu$m and [SIII]\,18.71\,$\mu$m lines, for which our predictions are substantially higher than those from all the other models, the [NeII]\,12.81\,$\mu$m line, for which our predictions are considerably lower in the redshift bin $1.25<z<1.75$ and higher at larger redshifts, and the [NeIII]\,15.55\,$\mu$m and H$_{2}$\,17.03\,$\mu$m lines in some redshift bins. Apart from these cases, most differences are within a factor $\sim 2$.

In Table~\ref{tab:counts_spica_tot} the comparison with \citet{Spin12} is made in terms of the total number of $\ge 5\,\sigma$ detections. Again, differences are generally within a factor of 2, but can be much larger, in particular, again, for the PAH\,11.25\,$\mu$m and [SIII]\,18.71\,$\mu$m lines.

\begin{figure*}
\makebox[\textwidth][c]{
\includegraphics[trim=0.5cm 2.8cm 1.0cm 3.3cm,clip=true,width=1.1\textwidth, angle=0]{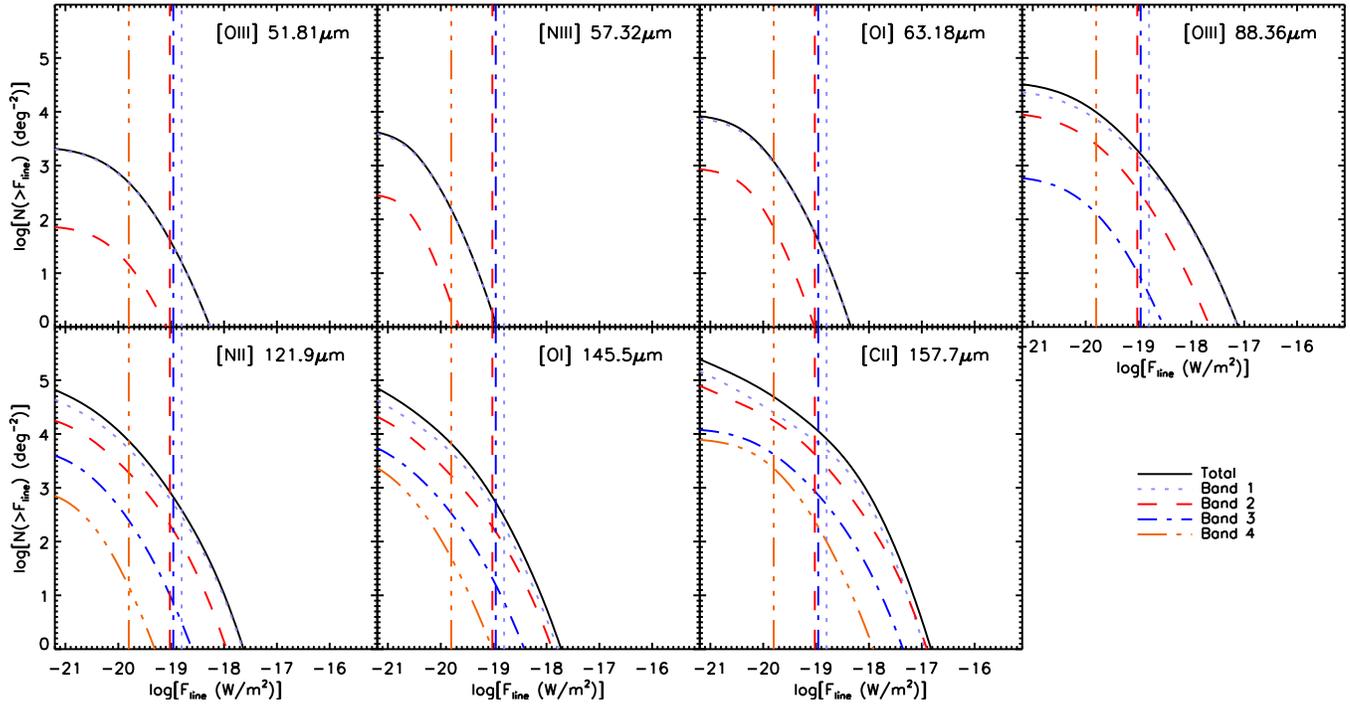}
}
\caption{Integral counts of star forming galaxies as a function of line fluxes over the full CCAT wavelength range (\emph{black solid line}) and in each of its four bands. The \emph{vertical lines} correspond to the detection limits in the four bands for 1\,hour integration per FoV.}
 \label{fig:intcounts_ccat_all}
\end{figure*}

In the case of CCAT, the $5\sigma$ detection limits for an integration of 1\,hour per FoV (reference detection limits) are \citep{Glenn2012} $1.6\times10^{-19}\,\hbox{W}/\hbox{m}^{2}$ for the first band ($\,312.3-380.9\,\mu$m), $9.5\times10^{-20}\,\hbox{W}/\hbox{m}^{2}$ for the second band ($\,416.4-499.7\,\mu$m), $1.1\times10^{-19}\,\hbox{W}/\hbox{m}^{2}$ for the third band ($\,587.8-648.9\,\mu$m) and $1.6\times10^{-20}\,\hbox{W}/\hbox{m}^{2}$ for the fourth band ($\,803.7-1102.2\,\mu$m). The integral counts in the CCAT wavelength range for the 7 lines within the SPICA/SAFARI range, are shown in Fig.~\ref{fig:intcounts_ccat_all}.

The slopes of the CCAT integral counts just below the detection limits of the reference detection limits for the bands giving the largest contributions  are $< 2$  for 5 out of 7 of the considered lines (the 2 exceptions are the [NIII]\,57.32\,$\mu$m and [OI]\,63.18\,$\mu$m lines), implying that the number of detections for given observing time increases more with the surveyed area than with survey depth. The predicted redshift distributions of galaxies detectable over $1\,\hbox{deg}^2$ with 1\,h integration per FoV in each of the 7 emission lines are presented in Table~\ref{tab:counts_ccat}. The predictions range from 14 to more than $12,000$ (in the [CII]157.7$\mu$m line) sources. In the latter line, such CCAT survey will be able to detect, in the redshift interval $5<z<6$, about 490 unlensed and 12 strongly lensed proto-spheroidal galaxies. Some 12 galaxies in the range $6<z<8$ should be detected in other lines. Thus a CCAT survey of a few hours will provide a good grasp of the early dust-obscured phase of galaxy formation up to the re-ionization epoch. 

\begin{figure*}
\hspace{+0.0cm}
\includegraphics[trim=0.1cm 2.2cm 0.75cm 2.6cm,clip=true,width=0.99\textwidth, angle=0]{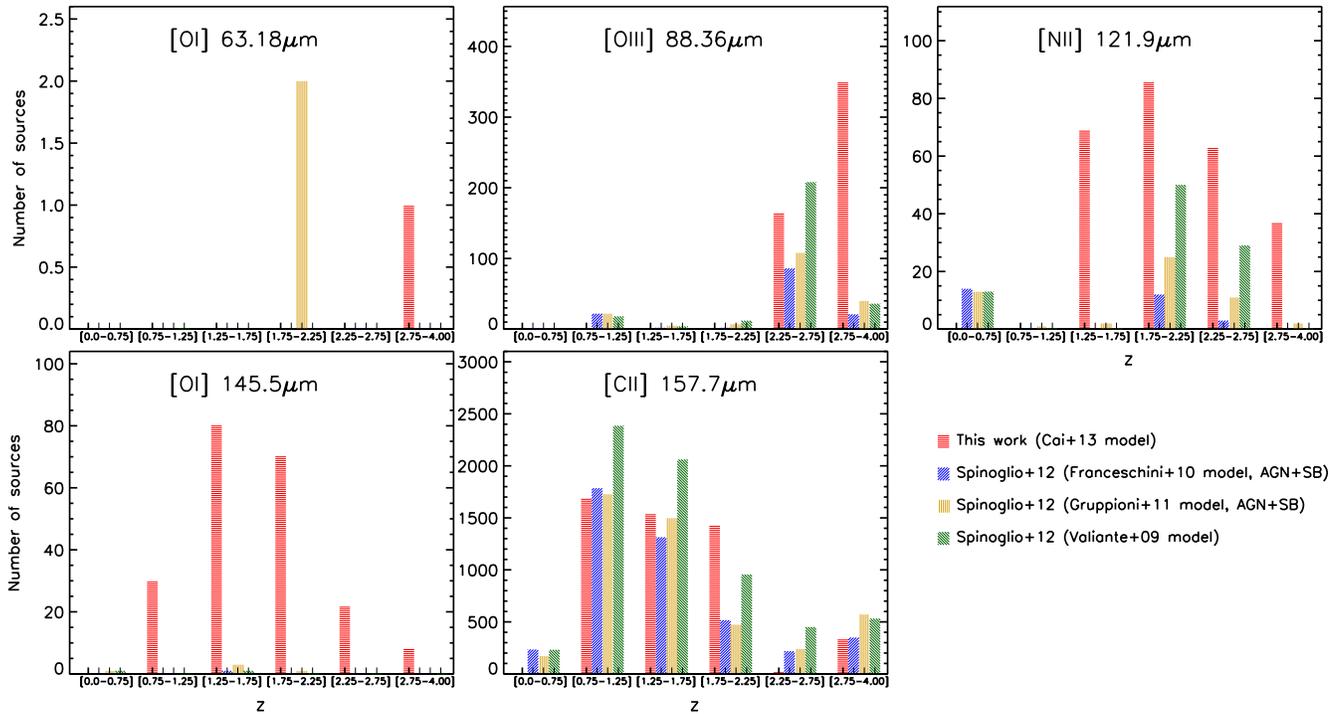}
\vspace{+0.0cm}
\caption{Redshift distributions of galaxies detectable in 5 lines by the CCAT survey considered by \citet[][exposure time: 1\,h per FoV; area: $0.5\,\hbox{deg}^2$]{Spin12}. The predictions of our model (\emph{bars with red horizontal dashes}) are compared with those of the 3 models used by \citet{Spin12}; see the legend on the bottom right.}
 \label{fig:histoCCAT}
\end{figure*}

Figure~\ref{fig:histoCCAT} shows that the redshift distributions predicted by our model for the CCAT survey considered by \citet[][exposure time: 1\,h per FoV; area: $0.5\,\hbox{deg}^2$]{Spin12} in the 5 lines for which results are also presented by \citet{Spin12} are widely different from those implied by the 3 models used in the latter paper. The differences, arising from those in the model luminosity functions at $z\geq3$ and in the relationships between line and IR luminosities, are much larger than found for the SPICA/SAFARI reference survey, except for the [CII]157.7$\mu$m line. These large differences are reflected in the comparison in terms of the total number of detections by the reference survey (Table~\ref{tab:counts_ccat_tot}).

%

\section{Survey strategy}\label{sect:survey}

Table~\ref{tab:counts_spica2} shows how the number of SPICA/SAFARI detections is maximized at fixed observing time (450\,hours). The number of detections steadily decreases with increasing integration time per FoV (and correspondingly decreasing the surveyed area), except for a couple of lines ([NeIII]15.55\,$\mu$m and H$_{2}$17.03\,$\mu$m) for which however the number of detections is only weakly varying. This is because, as noted above, the slope of the counts below the detection limits of the reference survey is relatively flat. Going deeper is only marginally advantageous even for the detection of the highest $z$ galaxies (see, for example, the variation with survey depth of the number of objects at $z>3$, given in parenthesis).  This result is in agreement with that obtained by \citet{Neg13} for detections in 2 or more lines (see their Figure 2).

\begin{figure*}
\makebox[\textwidth][c]{
\includegraphics[trim=2.4cm 1.1cm 1.1cm 1.1cm,clip=true,width=0.48\textwidth, angle=0]{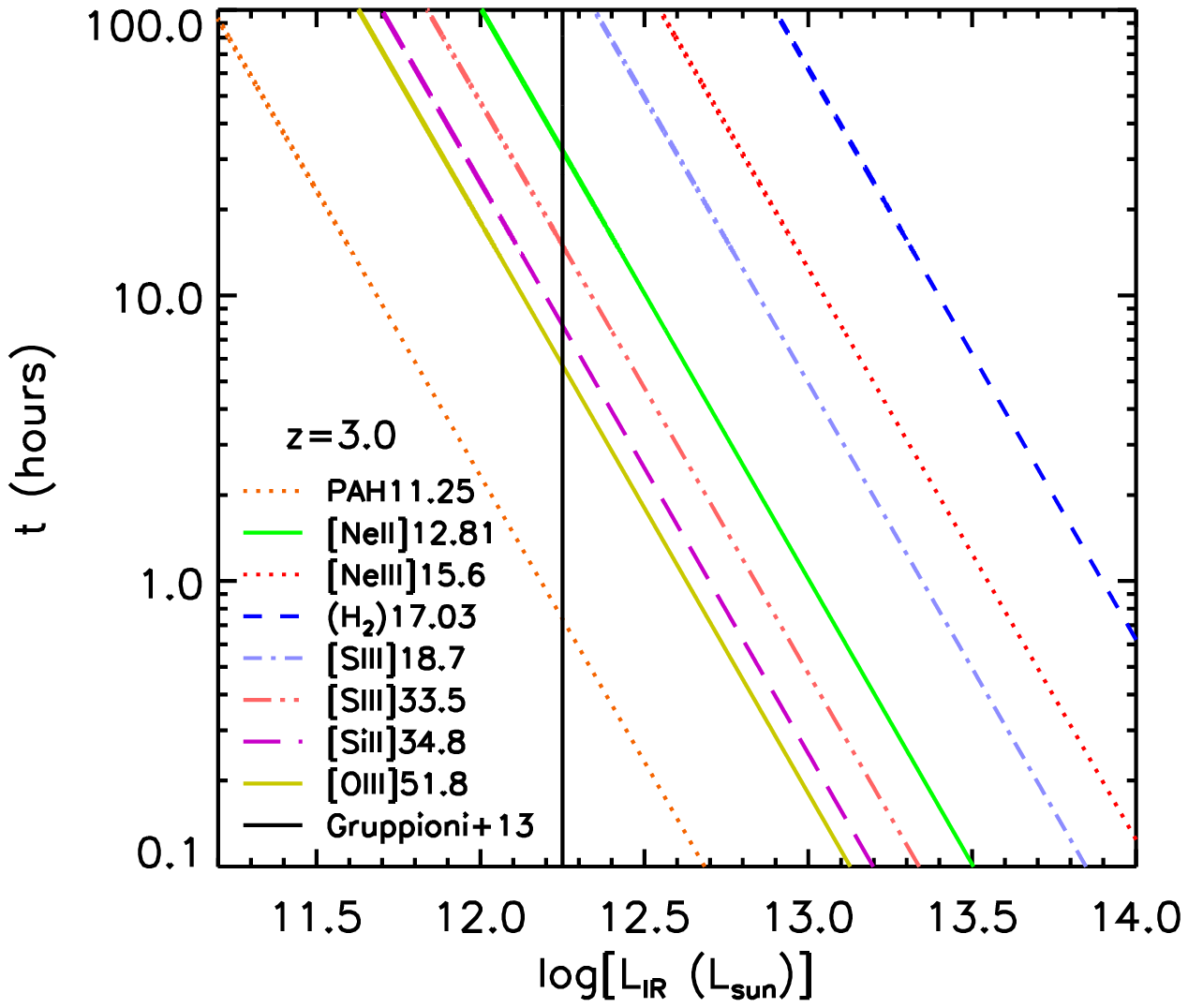}
\includegraphics[trim=2.4cm 1.1cm 1.1cm 1.1cm,clip=true,width=0.48\textwidth, angle=0]{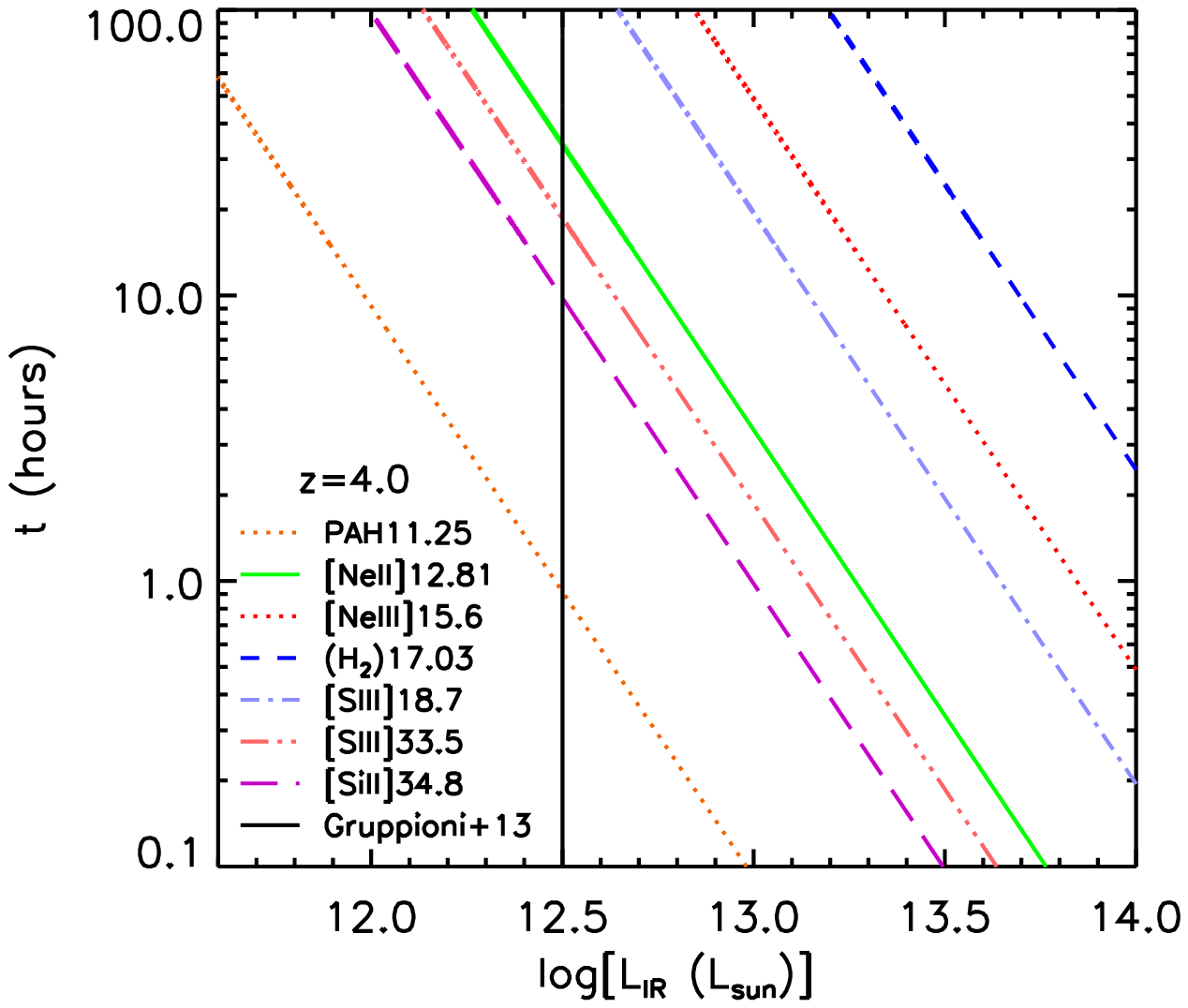}
}
\caption{SPICA/SAFARI exposure time per FoV required for a $5\sigma$ line detection (see legend inside each panel) as a function of IR luminosity for galaxies at $z=3$ (left) and $z=4$ (right). The \emph{vertical solid lines} correspond to the minimum luminosities represented in the IR luminosity functions determined by \citet{Grupp13} on the basis of \textit{Herschel}/PACS and SPIRE  surveys.}
 \label{fig:strategy}
\end{figure*}

The statistics of objects at $z\ge 3$ detected by the SPICA/SAFARI reference survey is quite poor. So, in order to investigate the star formation activity at high-$z$, it is expedient to complement the blind spectroscopic survey with follow-up observations of
the high-$z$ galaxies already discovered at (sub-)mm wavelengths over much larger areas. Figure~\ref{fig:strategy} shows, as an example, the SPICA/SAFARI exposure time per FoV necessary for a $5\sigma$ detection of lines detectable at $z=3$ and $z=4$ as a function of the IR luminosity. At these redshifts an exposure time of 10\,h will allow the detection in two lines of all the galaxies detected in the \textit{Herschel}/PACS and SPIRE surveys used by \citet{Grupp13} to derive the IR luminosity functions. A similar conclusion also holds at $z=2$. Note that only a small fraction of galaxies at $z\ge 2$ detected in these surveys have spectroscopic redshifts. Moreover, when SPT and \textit{Herschel} survey data will be fully available we will have samples of many hundreds of bright strongly lensed galaxies \citep[apparent luminosities $L_{\rm IR}>10^{13}\,L_{\odot}$;][]{Neg10,Vieira13}, that can be detected in few minutes in several lines by SPICA/SAFARI.

Also for CCAT, at fixed observing time, it is generally more advantageous to cover larger areas than to go deeper, even to detect high-$z$ galaxies, as shown by Table~\ref{tab:counts-ccat2}. The exception is the [NIII]57.32\,$\mu$m line, whose counts keep a slope larger than 2 down to fluxes corresponding to exposure times of a few hundred hours.

A CCAT survey of 1000\,hours, covering $1000\,\hbox{deg}^2$ in 1\,hour integration, will be able to detect $1.3\times10^{7}$ ($2.9\times10^{6}$ at $z>3$) star forming galaxies in at least one spectral line, $9.1\times10^{5}$ ($2.6\times10^{5}$ at $z>3$) in at least two lines and $8.8\times10^{4}$ ($2.6\times10^{4}$ at $z>3$) in at least three lines. The number of strongly lensed proto-spheroidal galaxies detectable by the same CCAT survey in at least one line is $1.1\times10^{5}$ ($6.7\times10^{4}$ at $z>3$), in at least two lines $2.0\times10^{4}$ ($1.3\times10^{4}$ at $z>3$) and in at least three lines $4110$ ($2750$ at $z>3$).


\begin{table*}
\centering
\footnotesize
\begin{tabular}{lcccc}
\hline
\hline
Spectral line & $t=1\,\hbox{h}$ & $t=10\,\hbox{h}$ & $t=100\,\hbox{h}$ & $t=1000\,\hbox{h}$ \\
  & $1000\,\hbox{deg}^2$ & $100\,\hbox{deg}^2$ & $10\,\hbox{deg}^2$ & $1\,\hbox{deg}^2$ \\
\hline
{\rm [OIII]51.81}$\mu$m &     $1.63\times10^{4}$ ($1.63\times10^{4}$) &        $1.18\times10^{4}$ ($1.18\times10^{4}$) &             $4819 (4819)$ &             $1140 (1140)$ \\
{\rm [NIII]57.32}$\mu$m &     $76 (76)$                               &        $992 (992)$ &                                         $1541 (1541)$ &             $1063 (1063)$ \\
{\rm [OI]63.18}$\mu$m &       $1.45\times10^{4}$ ($1.45\times10^{4}$) &         $2.04\times10^{4}$ ($2.04\times10^{4}$) &          $1.26\times10^{4}$ ($1.26\times10^{4}$) &           $4011$ ($4011$) \\
{\rm [OIII]88.36}$\mu$m &     $1.23\times10^{6}$ ($6.26\times10^{5}$) &       $4.20\times10^{5}$ ($2.29\times10^{5}$) &      $1.08\times10^{5}$ ($5.98\times10^{4}$) &         $2.01\times10^{4}$ ($1.10\times10^{4}$) \\
{\rm [NII]121.9}$\mu$m &      $5.27\times10^{5}$ ($3.29\times10^{4}$) &       $2.57\times10^{5}$ ($2.57\times10^{4}$) &        $9.16\times10^{4}$ ($1.13\times10^{4}$) &           $2.41\times10^{4}$ ($3054$) \\
{\rm [OI]145.5}$\mu$m &       $4.68\times10^{5}$ ($6.55\times10^{4}$) &        $2.56\times10^{5}$ ($4.33\times10^{4}$) &       $9.16\times10^{4}$ ($1.66\times10^{4}$) &           $2.40\times10^{4}$ ($4270$) \\
{\rm [CII]157.7}$\mu$m &      $1.23\times10^{7}$ ($2.42\times10^{6}$) &       $3.04\times10^{6}$ ($5.39\times10^{5}$) &   $6.09\times10^{5}$ ($8.19\times10^{4}$) &        $1.09\times10^{5}$ ($1.00\times10^{4}$) \\
\hline
\hline
\end{tabular}
\caption{Numbers of star forming galaxies detectable by CCAT surveys of 1000\,h covering areas from $1000\,\hbox{deg}^2$ to $1\,\hbox{deg}^2$ with integration times per FoV correspondingly ranging from 1 to 1000\,h, assuming the goal FoV of $1\,\hbox{deg}^2$. The numbers of galaxies at $z>3$ are given in parenthesis.}
\label{tab:counts-ccat2}
\end{table*}

%

%

%
\section{Conclusions}\label{sect:conclusions}

Our understanding of the cosmological evolution of IR galaxies has dramatically improved in recent years. Observational determinations of their IR luminosity functions up to $z\simeq 4$ are now available. A major step forward will be the characterization of their  physical properties, such as the intensity of their interstellar radiation fields,  their chemical abundances, the temperatures and densities of their ISM. This will be made possible by the IR spectroscopy provided by forthcoming facilities and in particular by the SAFARI instrument on SPICA and by CCAT. To optimize the survey strategy and to understand the redshift and galaxy luminosity ranges that can be measured it is necessary to have detailed predictions of the line luminosity functions as a function of redshift and of the corresponding number counts as a function of line flux.

A key ingredient to make such predictions is the relationship between line and total IR luminosity, that will allow us to take advantage of the wealth of information on the evolution of the IR luminosity function. A difficulty in this respect is that reliable measurements of both IR continuum and line luminosities are available for limited and generally heterogeneous samples of galaxies. To account for possible biases in these sample due, e.g., to variable attenuation of the lines by dust, we have carried out extensive simulations exploiting the public library of line luminosities compiled by \citet{Panuzzo2003}. This library provides line luminosities for ranges  of gas density, gas metallicity and gas filling factor fully adequate to describe the emission properties of the majority of star forming galaxies. To estimate the distribution of the line-to-IR luminosity ratios we first simulated the spectral SEDs of dust-free star forming galaxies and then added in a self-consistent way the effect of dust extinction distributed over a realistic range. The results are consistent with line luminosities being proportional to the total IR luminosities. An exception to this proportionality is observationally found in the case of low-$z$ ULIRGs for lines originated outside the stellar birth clouds. In these objects such lines are apparently uncorrelated with $L_{\rm IR}$ and are fainter than expected from the correlation holding for the other sources. In this case, we have adopted a Gaussian distribution of line luminosities around the global mean value.

The distributions of line luminosities have then been exploited to estimate the redshift dependent line luminosity functions starting from the IR continuum luminosity functions. Again, simulations have been made to take fully into account the effect of the dispersions in the line to continuum luminosity ratios. The effect of the dispersion was found to be strong because of the steepness of the bright portion of the IR luminosity function.

For SPICA/SAFARI, we have worked out predictions for the source counts in 14 mid/far-IR emission lines associated to star formation and discussed the expected outcome of a survey covering $0.5\,\hbox{deg}^2$ with a 1\,hr integration/FoV (`the reference survey'). We find that the number of detections in a single line ranges from a few to thousands. In total we expect that more than 21,000 lines will be detected. About 4600, 2800, 1800, 1100 and 700 sources will be detected in at least 2, 3, 4, 5 and 6 lines, respectively. This means that the number of spectroscopically detected individual galaxies is $\simeq 9700$. Sources detected in at least one line include $\simeq 40$ strongly lensed galaxies at $z>1$.

The slopes of the predicted integral counts below the detection limits of the SPICA/SAFARI reference survey are found to be $<2$ for most (11 out of 14) of the considered lines. Thus, should more survey time be available, extending the area will produce more detections than going deeper. The reference survey is found to maximize both the number of total detections and of detections at $z>3$, compared to deeper surveys, at fixed observing time (450\,hours).

Follow-up SPICA/SAFARI observations of about 10\,h per FoV will detect in two lines essentially all the $z\ge 2$ galaxies discovered by the \textit{Herschel}/PACS and SPIRE surveys used by \citet{Grupp13} to build their redshift-dependent IR luminosity functions. Only a small fraction of these galaxies have spectroscopic redshifts. 

We have also presented predictions, for the 7 longer wavelength lines in our set, of the number detections by a CCAT survey of 1000 hours, covering areas ranging from 1 to $1000\,\hbox{deg}^2$ with exposure times per FoV (assumed to be $1\,\hbox{deg}^2$) correspondingly ranging from 1 to 1000\,h. As in the SPICA/SAFARI case, we find that, at fixed observing time, it is more advantageous to cover larger areas than to go deeper. The number of detected galaxies in a survey of 1000\,h with an exposure of 1\,h per FoV ranges from some tens to more than ten million. In particular we expect $\simeq 10^7$ detections in the [CII]157.7$\mu$m line, including $\simeq 2.4\times 10^6$ galaxies at $z>3$ and about $4.9\times 10^5$ unlensed and $1.2\times 10^4$ strongly lensed galaxies at $z>5$. 

A comparison with the predicted numbers of SPICA/SAFARI detections for the reference survey worked out by \citet{Spin12} based on 3 models, substantially different from the one used here, shows that differences are in most cases within a factor of two, although occasionally are much larger. The discrepancies with the \citet{Spin12} predictions are more substantial in the case of CCAT, except in the case of the [CII]157.7\,$\mu$m line.


\section*{Acknowledgements} We are grateful to the anonymous referee for many constructive comments that helped us improving this paper and to Pasquale Panuzzo for clarifications on the use of his emission line library. We acknowledge financial support from ASI/INAF Agreement I/072/09/0 for the {\it Planck} LFI activity of Phase E2, from the PRIN MIUR 2009 ``Millimeter and sub-millimeter spectroscopy for high resolution studies of primeval galaxies and clusters of galaxies'' and from PRIN INAF 2012, project ``Looking into the dust-obscured phase of galaxy formation through cosmic zoom lenses in the Herschel Astrophysical Large Area Survey''.


\begin{thebibliography}{}
%
\bibitem[\protect\citeauthoryear{Amblard et
al.}{2011}]{Amblard11} Amblard A., et al., 2011, Natur, 470, 510
%
\bibitem[\protect\citeauthoryear{Baugh et al.}{2005}]{Baugh05}
Baugh C.~M., Lacey C.~G., Frenk C.~S., Granato G.~L., Silva L., Bressan A.,
Benson A.~J., Cole S., 2005, MNRAS, 356, 1191
%
\bibitem[\protect\citeauthoryear{Bernardi et al.}{2010}]{Bernardi2010} Bernardi M., Shankar F., Hyde J.~B., Mei S., Marulli F., Sheth R.~K., 2010, MNRAS, 404, 2087
%
\bibitem[\protect\citeauthoryear{Bernard-Salas et al.}{2009}]{Bern09} Bernard-Salas J., et al., 2009, ApJS, 184, 230
%
\bibitem[\protect\citeauthoryear{Bertelli et
al.}{1994}]{Bertelli1994} Bertelli G., Bressan A., Chiosi C., Fagotto F., Nasi E., 1994, A\&AS, 106, 275
%
%
\bibitem[\protect\citeauthoryear{Brandl et al.}{2006}]{Brandl06}
Brandl B.~R., et al., 2006, ApJ, 653, 1129
%
\bibitem[\protect\citeauthoryear{Brandl et al.}{2009}]{Brandl09}
Brandl B.~R., et al., 2009, ApJ, 699, 1982
%
\bibitem[\protect\citeauthoryear{Brauher, Dale,
\& Helou}{2008}]{Brauher08} Brauher J.~R., Dale D.~A., Helou G., 2008, ApJS, 178, 280
%
\bibitem[\protect\citeauthoryear{Bressan, Chiosi,
\& Fagotto}{1994}]{Bressan94} Bressan A., Chiosi C., Fagotto F., 1994, ApJS, 94, 63
%
%
\bibitem[\protect\citeauthoryear{Bruzual
\& Charlot}{2003}]{BC03} Bruzual G., Charlot S., 2003, MNRAS, 344, 1000
%
\bibitem[\protect\citeauthoryear{Cai et al.}{2013}]{Cai13}
Cai Z.-Y., et al., 2013, ApJ, 768, 21
%
\bibitem[\protect\citeauthoryear{Carilli
\& Walter}{2013}]{CarilliWalter2013} Carilli C., Walter F., 2013, arXiv, arXiv:1301.0371
%
\bibitem[\protect\citeauthoryear{Carral et al.}{1994}]{Carral94}
Carral P., Hollenbach D.~J., Lord S.~D., Colgan S.~W.~J., Haas M.~R., Rubin
R.~H., Erickson E.~F., 1994, ApJ, 423, 223
%
\bibitem[\protect\citeauthoryear{Chabrier}{2003}]{Chabrier2003}
Chabrier G., 2003, PASP, 115, 763
%
\bibitem[\protect\citeauthoryear{Charlot
\& Fall}{2000}]{CF00} Charlot S., Fall S.~M., 2000, ApJ, 539, 718
%
%
\bibitem[\protect\citeauthoryear{Colbert et al.}{1999}]{Colbert99} Colbert J.~W., et al., 1999, ApJ, 511, 721
%
\bibitem[\protect\citeauthoryear{Cooray et
al.}{2010}]{Cooray10} Cooray A., et al., 2010, A\&A, 518, L22
%
\bibitem[\protect\citeauthoryear{Coppin et al.}{2012}]{Coppin12}
Coppin K.~E.~K., et al., 2012, MNRAS, 427, 520
%
\bibitem[\protect\citeauthoryear{Cormier et
al.}{2012}]{Cormier12} Cormier D., et al., 2012, A\&A, 548, A20
%
\bibitem[\protect\citeauthoryear{Cox et al.}{2011}]{Cox11}
Cox P., et al., 2011, ApJ, 740, 63
%
\bibitem[\protect\citeauthoryear{da Cunha, Charlot,
\& Elbaz}{2008}]{daCunha08} da Cunha E., Charlot S., Elbaz D., 2008, MNRAS, 388, 1595
%
\bibitem[\protect\citeauthoryear{da Cunha et
al.}{2010}]{daCunha10} da Cunha E., Charmandaris V., D{\'{\i}}az-Santos T., Armus L., Marshall J.~A., Elbaz D., 2010, A\&A, 523, A78
%
%
\bibitem[\protect\citeauthoryear{Dav{\'e} et
al.}{2010}]{Dave10} Dav{\'e} R., Finlator K., Oppenheimer
B.~D., Fardal M., Katz N., Kere{\v s} D., Weinberg D.~H., 2010, MNRAS, 404,
1355
%
\bibitem[\protect\citeauthoryear{De Breuck et
al.}{2011}]{DeBreuck11} De Breuck C., Maiolino R., Caselli P., Coppin K., Hailey-Dunsheath S., Nagao T., 2011, A\&A, 530, L8
%
\bibitem[\protect\citeauthoryear{Devost et al.}{2004}]{Devost04}
Devost D., et al., 2004, ApJS, 154, 242
%
\bibitem[\protect\citeauthoryear{Farrah et al.}{2007}]{Farrah07}
Farrah D., et al., 2007, ApJ, 667, 149
%
\bibitem[\protect\citeauthoryear{Ferland}{2001}]{Ferland01} Ferland G.~L. 2001, Hazy, a brief introduction to CLOUDY, University of Kentucky, Department of Physics and Astronomy Internal Report
%
\bibitem[\protect\citeauthoryear{Fiolet et
al.}{2010}]{Fiolet10} Fiolet N., et al., 2010, A\&A, 524, A33
%
\bibitem[\protect\citeauthoryear{Fischer et
al.}{2010}]{Fischer2010} Fischer J., et al., 2010, A\&A, 518, L41
%
\bibitem[\protect\citeauthoryear{Franceschini et
al.}{2010}]{Franc10} Franceschini A., Rodighiero G., Vaccari M., Berta S., Marchetti L., Mainetti G., 2010, A\&A, 517, A74
%
\bibitem[\protect\citeauthoryear{Fu et al.}{2013}]{Fu13} Fu
H., et al., 2013, Natur, 498, 338
%
\bibitem[\protect\citeauthoryear{Galliano}{2006}]{Galliano06}
Galliano F., 2006, astro, arXiv:astro-ph/0610852
%
\bibitem[\protect\citeauthoryear{George et al.}{2013}]{George2013} George R.~D., et al., 2013, MNRAS, in press

\bibitem[\protect\citeauthoryear{Glenn et al.}{2012}]{Glenn2012} Glenn J., et al., 2012, http://www.ccatobservatory.org/docs/ pdfs/CCAT\_Science\_Requirements\_R1.pdf

\bibitem[\protect\citeauthoryear{Gonz{\'a}lez-Nuevo et al.}{2012}]{GonzalezNuevo2012} Gonz{\'a}lez-Nuevo J., et al., 2012, ApJ,
749, 65

\bibitem[\protect\citeauthoryear{Graci{\'a}-Carpio et
al.}{2008}]{GraciaCarpio2008} Graci{\'a}-Carpio J., Garc{\'{\i}}a-Burillo S., Planesas P., Fuente A., Usero A., 2008, A\&A, 479, 703
%
\bibitem[\protect\citeauthoryear{Graci{\'a}-Carpio et
al.}{2011}]{Gracia11} Graci{\'a}-Carpio J., et al., 2011, ApJ,
728, L7
%
\bibitem[\protect\citeauthoryear{Granato et
al.}{2004}]{Granato2004} Granato G.~L., De Zotti G., Silva L.,
Bressan A., Danese L., 2004, ApJ, 600, 580
%
\bibitem[\protect\citeauthoryear{Gruppioni et
al.}{2011}]{Grupp11} Gruppioni C., Pozzi F., Zamorani G.,
Vignali C., 2011, MNRAS, 416, 70
%
\bibitem[\protect\citeauthoryear{Gruppioni et al.}{2013}]{Grupp13} Gruppioni C., et al., 2013, MNRAS, 432, 23
%
\bibitem[\protect\citeauthoryear{Hailey-Dunsheath et
al.}{2010}]{Hailey-Dunsheath10} Hailey-Dunsheath S., Nikola T., Stacey
G.~J., Oberst T.~E., Parshley S.~C., Benford D.~J., Staguhn J.~G., Tucker
C.~E., 2010, ApJ, 714, L162
%
\bibitem[\protect\citeauthoryear{Higdon et al.}{2006}]{Higdon06}
Higdon S.~J.~U., Armus L., Higdon J.~L., Soifer B.~T., Spoon H.~W.~W.,
2006, ApJ, 648, 323
\bibitem[\protect\citeauthoryear{Imanishi et
al.}{2007}]{Imanishi07} Imanishi M., Dudley C.~C., Maiolino R.,
Maloney P.~R., Nakagawa T., Risaliti G., 2007, ApJS, 171, 72
%
\bibitem[\protect\citeauthoryear{Imanishi}{2009}]{Imanishi09}
Imanishi M., 2009, ApJ, 694, 751
%
\bibitem[\protect\citeauthoryear{Imanishi, Maiolino,
\& Nakagawa}{2010}]{Imanishi10} Imanishi M., Maiolino R., Nakagawa T., 2010, ApJ, 709, 801
%
\bibitem[\protect\citeauthoryear{Ivison et
al.}{2010}]{Ivison10} Ivison R.~J., et al., 2010, A\&A, 518, L35
%
\bibitem[\protect\citeauthoryear{Ivison et al.}{2013}]{Ivison13}
Ivison R.~J., et al., 2013, ApJ, 772, 137
%
\bibitem[\protect\citeauthoryear{Karim et al.}{2013}]{Karim2013} Karim A., et al., 2013, MNRAS, 432, 2

\bibitem[\protect\citeauthoryear{Kaufman et
al.}{1999}]{Kaufman1999} Kaufman M.~J., Wolfire M.~G., Hollenbach
D.~J., Luhman M.~L., 1999, ApJ, 527, 795
%
\bibitem[\protect\citeauthoryear{Lacey et al.}{2010}]{Lacey10}
Lacey C.~G., Baugh C.~M., Frenk C.~S., Benson A.~J., Orsi A., Silva L.,
Granato G.~L., Bressan A., 2010, MNRAS, 405, 2
%
\bibitem[\protect\citeauthoryear{Lapi
\& Cavaliere}{2011}]{LC11} Lapi A., Cavaliere A., 2011, ApJ, 743, 127
%
\bibitem[\protect\citeauthoryear{Lapi et al.}{2006}]{Lapi06}
Lapi A., Shankar F., Mao J., Granato G.~L., Silva L., De Zotti G., Danese
L., 2006, ApJ, 650, 42
%
\bibitem[\protect\citeauthoryear{Lapi et al.}{2011}]{Lapi11}
Lapi A., et al., 2011, ApJ, 742, 24
%
\bibitem[\protect\citeauthoryear{Lapi et al.}{2012}]{Lapi12}
Lapi A., Negrello M., Gonz{\'a}lez-Nuevo J., Cai Z.-Y., De Zotti G., Danese
L., 2012, ApJ, 755, 46
%
\bibitem[\protect\citeauthoryear{Luhman et al.}{1998}]{Luhman98}
Luhman M.~L., et al., 1998, ApJ, 504, L11
%
\bibitem[\protect\citeauthoryear{Luhman et al.}{2003}]{Luhman2003}
Luhman M.~L., Satyapal S., Fischer J., Wolfire M.~G., Sturm E., Dudley
C.~C., Lutz D., Genzel R., 2003, ApJ, 594, 758
%
\bibitem[\protect\citeauthoryear{Maddox et al.}{2010}]{Maddox2010} Maddox S.~J., et al., 2010, A\&A, 518, L11

\bibitem[\protect\citeauthoryear{Maiolino et
al.}{2009}]{Maiolino2009} Maiolino R., Caselli P., Nagao T., Walmsley M., De Breuck C., Meneghetti M., 2009, A\&A, 500, L1
%
\bibitem[\protect\citeauthoryear{Malhotra et
al.}{2001}]{Malhotra01} Malhotra S., et al., 2001, ApJ, 561, 766
%
\bibitem[\protect\citeauthoryear{Mao et al.}{2007}]{Mao07}
Mao J., Lapi A., Granato G.~L., de Zotti G., Danese L., 2007, ApJ, 667, 655
%
\bibitem[\protect\citeauthoryear{Mocanu et al.}{2013}]{Mocanu2013}
Mocanu L.~M., et al., 2013, arXiv, arXiv:1306.3470
%
\bibitem[\protect\citeauthoryear{Negishi et
al.}{2001}]{Negishi01} Negishi T., Onaka T., Chan K.-W., Roellig T.~L., 2001, A\&A, 375, 566
%
\bibitem[\protect\citeauthoryear{Negrello et
al.}{2007}]{Neg07} Negrello M., Perrotta F.,
Gonz{\'a}lez-Nuevo J., Silva L., de Zotti G., Granato G.~L., Baccigalupi
C., Danese L., 2007, MNRAS, 377, 1557
%
\bibitem[\protect\citeauthoryear{Negrello et
al.}{2010}]{Neg10} Negrello M., et al., 2010, Sci, 330, 800
%
\bibitem[\protect\citeauthoryear{Negrello et
al.}{2013}]{Neg13} Negrello M., Bonato M., Cai Z.-Y., de Zotti G., 2013, Proceedings of the SPICA Science Conference 2013, 18-21 June, Tokyo, Japan, ``From Exoplanets to Distant Galaxies: SPICA's New Window on the Cool Universe''
%
\bibitem[\protect\citeauthoryear{O'Dowd et al.}{2009}]{O'Dowd09}
O'Dowd M.~J., et al., 2009, ApJ, 705, 885
%
\bibitem[\protect\citeauthoryear{O'Dowd et al.}{2011}]{O'Dowd11}
O'Dowd M.~J., et al., 2011, ApJ, 741, 79
%
\bibitem[\protect\citeauthoryear{Panuzzo et
al.}{2003}]{Panuzzo2003} Panuzzo P., Bressan A., Granato G.~L., Silva L., Danese L., 2003, A\&A, 409, 99
%
\bibitem[\protect\citeauthoryear{Pereira-Santaella et
al.}{2010}]{Pereira-Santaella10} Pereira-Santaella M., Alonso-Herrero A.,
Rieke G.~H., Colina L., D{\'{\i}}az-Santos T., Smith J.-D.~T.,
P{\'e}rez-Gonz{\'a}lez P.~G., Engelbracht C.~W., 2010, ApJS, 188, 447
%
\bibitem[\protect\citeauthoryear{Planck Collaboration}{2011}]{PlanckCollaboration2011} Planck Collaboration, 2011, A\&A, 536, A18

\bibitem[\protect\citeauthoryear{Planck Collaboration XVI}{2013}]{PlanckCollaborationXVI2013} Planck Collaboration XVI, 2013, arXiv:1303.5076
%
%
\bibitem[\protect\citeauthoryear{Pope et al.}{2008}]{Pope08}
Pope A., et al., 2008, ApJ, 675, 1171
%
\bibitem[\protect\citeauthoryear{Puget
\& Leger}{1989}]{Puget89} Puget J.~L., Leger A., 1989, ARA\&A, 27, 161
%
\bibitem[\protect\citeauthoryear{Renzini}{2006}]{Renzini2006} Renzini A., 2006, ARA\&A, 44, 141
%
\bibitem[\protect\citeauthoryear{Riechers et
al.}{2013}]{Riechers13} Riechers D.~A., et al., 2013, Natur, 496,
329
%
\bibitem[\protect\citeauthoryear{Roelfsema et
al.}{2012}]{Roelfsema12} Roelfsema P., et al., 2012, SPIE, 8442,
%
\bibitem[\protect\citeauthoryear{Roussel et
al.}{2006}]{Roussel06} Roussel H., et al., 2006, ApJ, 646, 841
%
\bibitem[\protect\citeauthoryear{Rubin et al.}{1994}]{Rubin94}
Rubin R.~H., Simpson J.~P., Lord S.~D., Colgan S.~W.~J., Erickson E.~F.,
Haas M.~R., 1994, ApJ, 420, 772
%
\bibitem[\protect\citeauthoryear{Sajina et al.}{2007}]{Sajina07}
Sajina A., Yan L., Armus L., Choi P., Fadda D., Helou G., Spoon H., 2007,
ApJ, 664, 713
%
\bibitem[\protect\citeauthoryear{Sanders \& Mirabel}{1996}]{SandersMirabel1996} Sanders D.~B., Mirabel I.~F., 1996, ARA\&A, 34, 749

\bibitem[\protect\citeauthoryear{Saunders et
al.}{1990}]{Saunders1990} Saunders W., Rowan-Robinson M., Lawrence
A., Efstathiou G., Kaiser N., Ellis R.~S., Frenk C.~S., 1990, MNRAS, 242,
318
%
\bibitem[\protect\citeauthoryear{Silva et al.}{1998}]{Silva98}
Silva L., Granato G.~L., Bressan A., Danese L., 1998, ApJ, 509, 103
%
\bibitem[\protect\citeauthoryear{Spinoglio et
al.}{2012}]{Spin12} Spinoglio L., Dasyra K.~M., Franceschini
A., Gruppioni C., Valiante E., Isaak K., 2012, ApJ, 745, 171
%
\bibitem[\protect\citeauthoryear{Spinoglio \& Malkan}{1992}]{Spin92} Spinoglio L., Malkan M.~A., 1992, ApJ, 399, 504
%
\bibitem[\protect\citeauthoryear{Stacey et al.}{2010}]{Stacey10}
Stacey G.~J., Hailey-Dunsheath S., Ferkinhoff C., Nikola T., Parshley
S.~C., Benford D.~J., Staguhn J.~G., Fiolet N., 2010, ApJ, 724, 957
%
%
\bibitem[\protect\citeauthoryear{Swinbank et
al.}{2012}]{Swinbank12} Swinbank A.~M., et al., 2012, MNRAS, 427,
1066
%

\bibitem[\protect\citeauthoryear{Thornley et al.}{2000}]{Thornley2000} Thornley M.~D., Schreiber N.~M.~F., Lutz D., Genzel R., Spoon H.~W.~W., Kunze D., Sternberg A., 2000, ApJ, 539, 641

\bibitem[\protect\citeauthoryear{Tielens}{2008}]{Tielens2008} Tielens A.~G.~G.~M., 2008, ARA\&A, 46, 289
%
\bibitem[\protect\citeauthoryear{Tommasin et al.}{2010}]{Tommasin2010} Tommasin S., Spinoglio L., Malkan M.~A., Fazio G., 2010, ApJ, 709, 1257
\bibitem[\protect\citeauthoryear{Unger et
al.}{2000}]{Unger00} Unger S.~J., et al., 2000, A\&A, 355, 885
%
\bibitem[\protect\citeauthoryear{Valiante et
al.}{2009}]{Valiante09} Valiante E., Lutz D., Sturm E., Genzel R.,
Chapin E.~L., 2009, ApJ, 701, 1814
%
\bibitem[\protect\citeauthoryear{Veilleux et
al.}{2009}]{Veilleux09} Veilleux S., et al., 2009, ApJS, 182, 628
%
\bibitem[\protect\citeauthoryear{Vega et
al.}{2008}]{Vega08} Vega O., Clemens M.~S., Bressan A., Granato G.~L., Silva L., Panuzzo P., 2008, A\&A, 484, 631
%
\bibitem[\protect\citeauthoryear{Vieira et al.}{2010}]{Vieira10}
Vieira J.~D., et al., 2010, ApJ, 719, 763
%
\bibitem[\protect\citeauthoryear{Vieira et al.}{2013}]{Vieira13}
Vieira J.~D., et al., 2013, Natur, 495, 344
%
\bibitem[\protect\citeauthoryear{Viero et al.}{2013}]{Viero13}
Viero M.~P., et al., 2013, ApJ, 772, 77
%
\bibitem[\protect\citeauthoryear{Wagg et
al.}{2010}]{Wagg10} Wagg J., Carilli C.~L., Wilner D.~J., Cox P., De Breuck C., Menten K., Riechers D.~A., Walter F., 2010, A\&A, 519, L1
%
\bibitem[\protect\citeauthoryear{Walter et al.}{2012}]{Walter12}
Walter F., et al., 2012, Natur, 486, 233
%
\bibitem[\protect\citeauthoryear{Wang et al.}{2011}]{Wang11}
Wang J., et al., 2011, MNRAS, 413, 1373
%
\bibitem[\protect\citeauthoryear{Wardlow et al.}{2013}]{Wardlow2013} Wardlow J.~L., et al., 2013, ApJ, 762, 59

\bibitem[\protect\citeauthoryear{Wei{\ss} et al.}{2009}]{Weiss2009} Wei{\ss} A., et al., 2009, ApJ, 707, 1201

\bibitem[\protect\citeauthoryear{Wei{\ss} et al.}{2013}]{Weiss13}  Wei{\ss} A., et al., 2013, ApJ in press (arXiv:1303.2726)

\bibitem[\protect\citeauthoryear{Woody et al.}{2012}]{Woody2012} Woody D., et al., 2012, SPIE, 8444

%
\bibitem[\protect\citeauthoryear{Xia et al.}{2012}]{Xia2012}
Xia J.-Q., Negrello M., Lapi A., De Zotti G., Danese L., Viel M., 2012,
MNRAS, 422, 1324
%
\bibitem[\protect\citeauthoryear{Yan et al.}{2005}]{Yan05}
Yan L., et al., 2005, ApJ, 628, 604
%
\bibitem[\protect\citeauthoryear{Yan et al.}{2007}]{Yan07}
Yan L., et al., 2007, ApJ, 658, 778
%
\bibitem[\protect\citeauthoryear{Zhao et al.}{2003}]{Zhao2003}
Zhao D.~H., Jing Y.~P., Mo H.~J., B{\"o}rner G., 2003, ApJ, 597, L9
%


\end{thebibliography}
\end{document}